\documentclass[preprint,12pt]{elsarticle}

\usepackage[a4paper]{geometry}													% von Matthias' Veröffentlichung; HK, 28.06.2017
\usepackage{amssymb}
\usepackage{caption}
\usepackage{amsmath}
\usepackage{color}																	% Nur während der Dokumenterstellung; HK; 28.06.2017
\usepackage{colortbl}																% Für Kolorierung von Tabellen; HK, Di, 08.05.2018 (http://www.kkittel.de/wiki/doku.php?id=tabellen:farbige_zellen)
\newcommand{\Tref}[0]{293.15K }	% Referenztemperatur, HK, Di, 28.01.2020
\newcommand{\JCA}[0]{430.9}	% JC-Kennwerte, HK, Do, 30.01.2020
\newcommand{\JCB}[0]{908.7}	% JC-Kennwerte, HK, Do, 30.01.2020
\newcommand{\JCC}[0]{0.00447}	% JC-Kennwerte, HK, Do, 30.01.2020
\newcommand{\JCm}[0]{0.8584}	% JC-Kennwerte, HK, Do, 30.01.2020
\newcommand{\JCmOhneSechshundertC}[0]{0.7361}	% JC-Kennwerte, HK, Mi, 22.04.2020
\newcommand{\JCn}[0]{0.3854}	% JC-Kennwerte, HK, Do, 30.01.2020
\newcommand{\JCepsref}[0]{10^{-3}}	% JC-Kennwerte, HK, Do, 30.01.2020; ModifHK: Mo, 03.08.2020
\newcommand{\JCCRII}[0]{0.35}	% JC-n -R²-Wert, HK, Mo, 02.03.2020
\newcommand{\JCmRII}[0]{0.83}	% JC-m -R²-Wert, HK, Mo, 02.03.2020
\newcommand{\JCmRIIOhneSechshundertC}[0]{0.80}	% JC-m -R²-Wert, HK, Mo, 02.03.2020
\newcommand{\JCAohneBruchdaten}[0]{430.9}	% JC-Kennwerte, HK, Do, 30.01.2020
\newcommand{\JCBohneBruchdaten}[0]{1605.7}	% JC-Kennwerte, HK, Do, 30.01.2020
\newcommand{\JCnohneBruchdaten}[0]{0.5829}	% JC-Kennwerte, HK, Do, 30.01.2020
\makeatletter
\def\ps@pprintTitle{%
    \let\@oddhead\@empty
	\let\@evenhead\@empty
	\def\@oddfoot{\reset@font\hfil\thepage\hfil}
	\let\@evenfoot\@oddfoot
}
\makeatother

\begin{document}

\begin{frontmatter}

%% Title, authors and addresses

%% use the tnoteref command within \title for footnotes;
%% use the tnotetext command for theassociated footnote;
%% use the fnref command within \author or \address for footnotes;
%% use the fntext command for theassociated footnote;
%% use the corref command within \author for corresponding author footnotes;
%% use the cortext command for theassociated footnote;
%% use the ead command for the email address,
%% and the form \ead[url] for the home page:
%\title{Title\tnoteref{label1}}
%\tnotetext[label1]{}
%\author{Name\corref{cor1}\fnref{label2}}
%\ead{email address}
%% \ead[url]{home page}
%% \fntext[label2]{}
%% \cortext[cor1]{}
%% \address{Address\fnref{label3}}
%% \fntext[label3]{}

\title{Johnson Cook Flow Stress Parameter for Free Cutting Steel 50SiB8}

%% use optional labels to link authors explicitly to addresses:
%% \author[label1,label2]{}
%% \address[label1]{}
%% \address[label2]{}
\author{Hagen Klippel*}
\ead{klippel@iwf.mavt.ethz.ch}
%\author{Marcel Gerstgrasser*, Darko Smolenicki*, Ezio Cadoni**, Hans Roelofs***, Prof. Dr. Konrad Wegener*}
\author{Marcel Gerstgrasser*, Darko Smolenicki*, Ezio Cadoni**, Hans Roelofs***, Konrad Wegener*}
\address{*Institute of Machine Tools and Manufacturing (IWF), Department of Mechanical and Process Engineering, ETH Z\"urich, Leonhardstrasse 21, 8092 Z\"uerich, Switzerland

** DynaMat Lab, University of Applied Sciences of Southern Switzerland, 6952 Canobbio, Switzerland

*** R\&D, Swiss Steel AG, Emmenweidstr. 90, 6020 Emmenbr\"ucke, Switzerland}

%\address{Institute of Machine Tools and Manufacturing (IWF), Department of Mechanical and Process Engineering, ETH Z\"urich, Leonhardstrasse 21, 8092 Z\"uerich, Switzerland}

\begin{abstract}
%% Text of abstract
The present publication deals with the material characterization of the free cutting steel 50SiB8 for numerical simulations. Quasi-static tensile tests as well as Split Hopkinson Tension Bar (SHTB) tests at various strain rates and temperatures are used to deduce the parameters for a Johnson-Cook flow stress model. These parameters are then verified against the SHTB-experiments within a finite element model (FEM) of the SHTB-test within ABAQUS\textsuperscript{\textcopyright}.

\end{abstract}

\begin{keyword}
%% keywords here, in the form: keyword \sep keyword

%% PACS codes here, in the form: \PACS code \sep code

%% MSC codes here, in the form: \MSC code \sep code
%% or \MSC[2008] code \sep code (2000 is the default)
Johnson-Cook, material parameter determination, constitutive model, isotropic hardening, flow stress model, free cutting steel, Split Hopkinson Tension Bar test
\end{keyword}

\end{frontmatter}

%% \linenumbers

%% main text
\section{Introduction}
\label{Kap:Introduction}

The relatively new material 50SiB8 \cite{Chabbi2017, Roelofs2017} was developed by Swiss Steel\textsuperscript{\textcopyright} as a lead free alternative for classical free cutting steels, e.g. 11SMnPb30 and 16MnCrS5Pb. The development became necessary as regulatory requirements have tightened and in future may ban vehicle components containing heavy metals, such as lead \cite{EU_Direktive_2011}. The idea is to exchange lead by graphite inclusions in order to keep the good machinability of free cuttings steels. As the data basis for numerical simulations of this material is rather small \cite{Smolenicki2017, Akbari2019, Gerstgrasser2020}, this report is initiated in support to \cite{Gerstgrasser2020}, where the results of quasi-static tensile tests and Split Hopkinson Tension Bar (SHTB) tests were used to derive a modified Johnson-Cook fracture strain model for the free cutting steel 50SiB8. Here, the same test results are used to deduce the parameters for a flow stress model according to Johnson and Cook \cite{JohnsonCook1983}. The model is commonly used to describe metal plasticity within machining simulations and is given as:

\begin{equation}
	\sigma_y = \underbrace{\left( A+B \cdot \varepsilon_{pl}^n\right)}_{\text{1st term}} \hspace{3mm} \underbrace{\left[ 1+C \cdot ln \frac{\dot\varepsilon_{pl}}{\dot\varepsilon^0_{pl}} \right]}_{\text{2nd term}} \hspace{3mm} \underbrace{\left[1- \left(\frac{T-T_{ref}}{T_f-T_{ref}} \right)^m \right]}_{\text{3rd term}}
	\label{Glg:JohnsonCook}
\end{equation}

with $A$, $B$, $C$, $m$ and $n$ being material parameters, $\varepsilon_{pl}$ the plastic strain, $\dot \varepsilon_{pl}$ the plastic strain rate and $T$ the current temperature. $T_f$ is the melting temperature, $T_{ref}$ is the reference temperature and $\dot \varepsilon_{pl}^0$ the reference plastic strain rate. The first two terms describe hardening due to plastic strain and plastic strain rate, respectively. The third term controls thermal softening upon increasing temperature. The approach used here for the derivation of its 5 material parameters mainly follows \cite{Meyer2001b, Boehme2007}.

\section{Experimental Procedure}

Quasi-static and dynamic tests were performed. The complete test matrix is given with table \ref{Tab:TestMatrix} while test details are provided in the subsequent sections \ref{Kap:Experimente_Quasistatisch} and \ref{Kap:Experimente_SHTB}.

\begin{table}[ht]
	\centering
	\begin{tabular}{| l | c | c | c | c | c | c | c |}
		\hline
		Temperature & $20^\circ C$ & $200^\circ C$ & $400^\circ C$ & $600^\circ C$ & $800^\circ C$ & Test specimen & Test\\
		Strain Rate & & & & & & & \\
		\hline
		$0.001s^{-1}$ & 3 & - & - & - & - & \o=6mm, figure \ref{Bild:Zugprobe_SwissSteel} & quasi-static (tensile)\\
		\hline
		$500s^{-1}$ & 4 & 3 & 3 & 3 & 3 & \o=3mm, figure \ref{Bild:STHB_Zeichnung_SUPSI} & dynamic (SHTB)\\
		\hline
		$900s^{-1}$ & 4 & 3 & 3 & 3 & 3 & \o=3mm, figure \ref{Bild:STHB_Zeichnung_SUPSI} & dynamic (SHTB)\\
		\hline
		$1700s^{-1}$ & 4 & - & - & - & - & \o=3mm, figure \ref{Bild:STHB_Zeichnung_SUPSI} & dynamic (SHTB)\\
		\hline
	\end{tabular} 
	\caption{Static and dynamic test matrix with number of tests}
	\label{Tab:TestMatrix}
\end{table}

\subsection{Quasi-Static Tensile Tests}
\label{Kap:Experimente_Quasistatisch}

\begin{figure}[htp]
	\center{
		\includegraphics[scale=0.4]{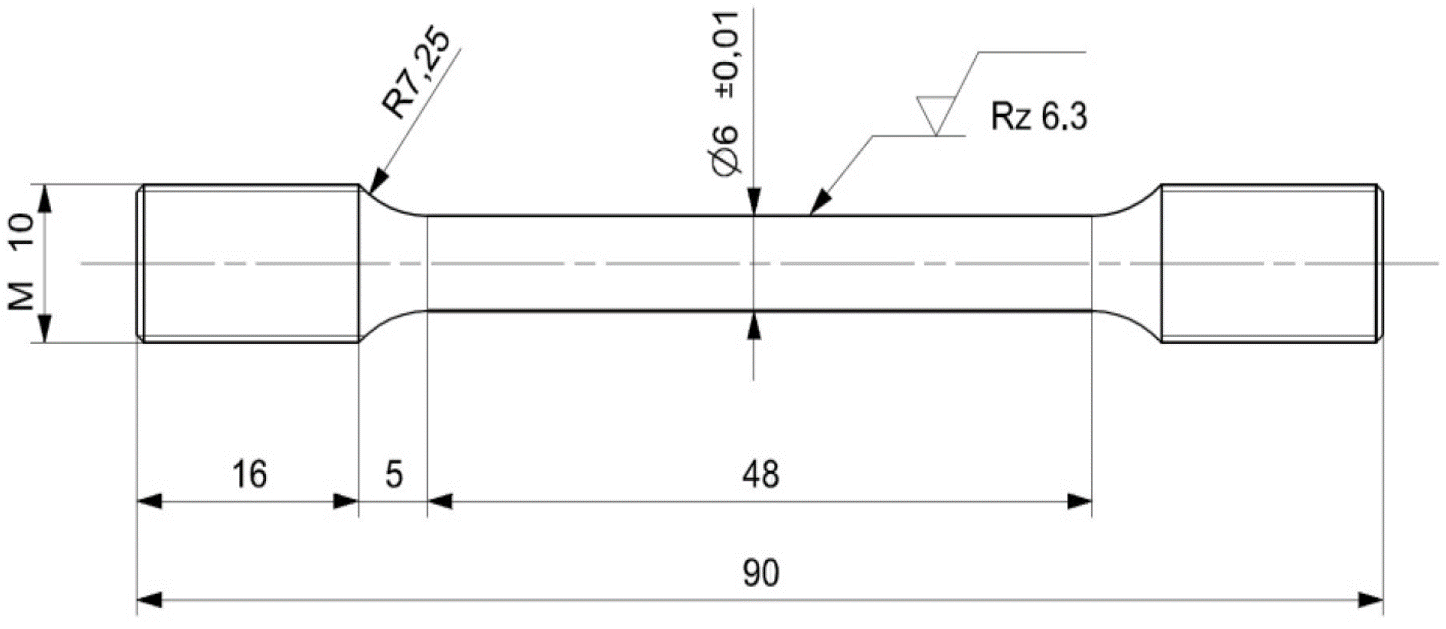}
	}
	\caption{Drawing of the quasi-static tensile test specimen according to \cite{Smolenicki2017}}
	\label{Bild:Zugprobe_SwissSteel}
\end{figure}

Three tensile tests were performed at room temperature (\Tref) at a very low strain rate of $\dot \varepsilon = 0,001/s$, see also table \ref{Tab:TestMatrix}. The low strain rate ensures almost quasi-static conditions. Unnotched test specimen were used with a diameter of 6mm. A drawing of the test specimen is shown in figure \ref{Bild:Zugprobe_SwissSteel}.% Three test specimen were loaded until rupture. The conducted tests are compiled in table \ref{Tab:StatischeTests}.

\begin{figure}[htp]
	\center{
		\includegraphics[scale=0.8]{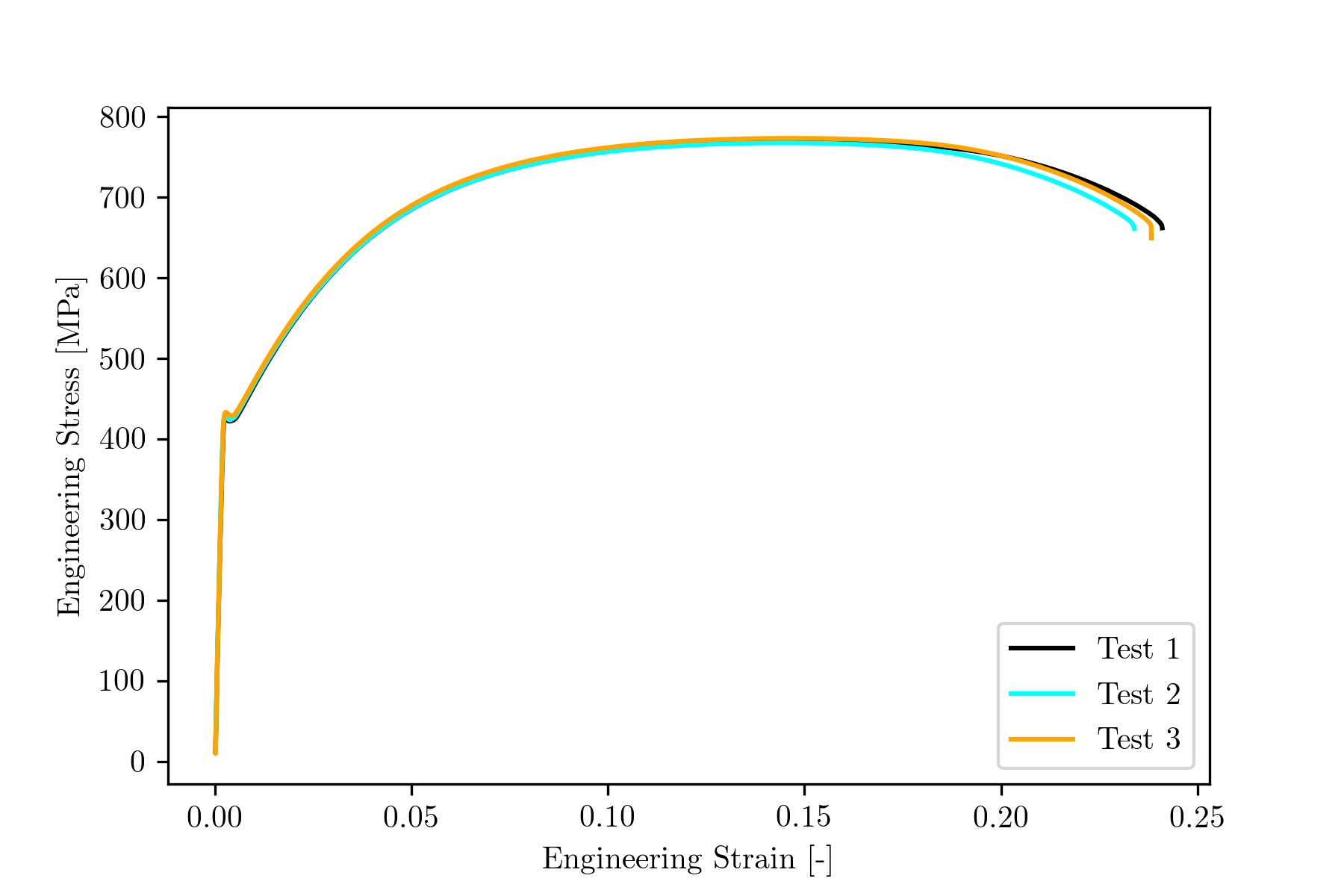}
	}
	\caption{Engineering stress-strain curves from quasi-static tests at room temperature}
	\label{Bild:Quasistatisch}
\end{figure}

The engineering stress-strain curves were recorded and are shown in figure \ref{Bild:Quasistatisch}. From these measurements the Johnson-Cook parameters A, B and n were fitted, see section \ref{Kap:ABn_Anpassung}.

\subsection{Testing at Higher Strain Rates}
\label{Kap:Experimente_SHTB}

\begin{figure}[htp]
	\center{
		\includegraphics[scale=0.7]{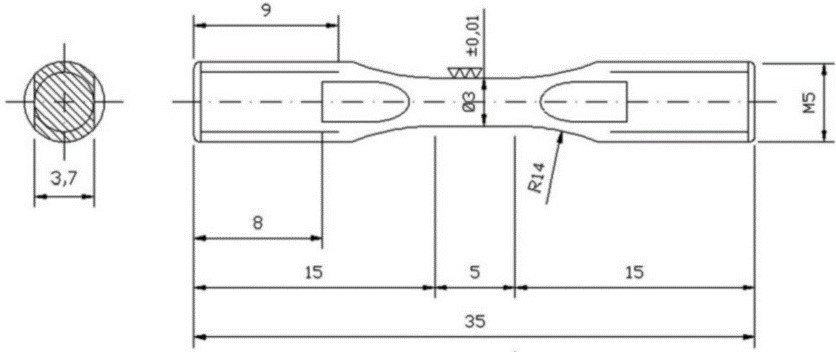}
	}
	\caption{Drawing of the SHTB test specimen according to \cite{Smolenicki2017}}
	\label{Bild:STHB_Zeichnung_SUPSI}
\end{figure}

Tests at higher strain rates were performed by means of the SHTB device with unnotched specimen and a diameter of 3mm. A drawing \cite{Smolenicki2017} of the SHTB test specimen is provided with figure \ref{Bild:STHB_Zeichnung_SUPSI}. After the tests for every SHTB test the average strain rate was evaluated between initial yield point and ultimate tensile strength \cite{Forni2016}.

More details about the test setup are provided in \cite{Smolenicki2017, Gerstgrasser2020}.
The results were used to identify:

\begin{itemize}
	\item the strain rate dependency (\textit{parameter C}) at room temperature \Tref for targeted strain rates of 500/s, 900/s and $\approx 1700/s$, each test with 4 repetitions and
	\item the temperature dependency (\textit{parameter m}) at room temperature and at elevated temperatures of $200^\circ C$, $400^\circ C$, $600^\circ C$ and $800^\circ C$ and targeted strain rates of 500/s and 900/s.
\end{itemize}

All conducted SHTB tests are compiled in table \ref{Tab:TestMatrix}.

\section{Material Parameter Determination}

\subsection{Assessment of True Stress and True Strain}

During the measurements engineering strain and stress values were recorded and were later converted into true strains and stresses. Until uniform elongation $A_G$ the conversion can be performed by the following equations given in \cite{Boehme2007}:

\begin{equation}
	\varepsilon_{true} = ln(1+\varepsilon_{eng})
	\label{Glg:WahreDehnung}
\end{equation}

and

\begin{equation}
	\sigma_{true} = \sigma_{eng} \cdot (1+\varepsilon_{eng})
	\label{Glg:WahreSpannung}
\end{equation}

The true strain in equation (\ref{Glg:WahreDehnung}) consists of elastic and plastic contributions. According to \cite{Boehme2007}, assuming an additive split of both components, the true plastic strain can be computed by:

\begin{equation}	% Gleichung (8.5) in Boehme2007
	\varepsilon_{true}^{pl} = \varepsilon_{true} - \varepsilon^{el}_{true} = \varepsilon_{true} - \frac{\sigma_{true}}{E}
	\label{Glg:WahrePlastischeDehnung}
\end{equation}

Beyond uniform elongation $A_G$ the conversions (\ref{Glg:WahreDehnung}) and (\ref{Glg:WahreSpannung}) are invalid. The determination of true stresses and strains would require ad-hoc tracking of the progressively reducing diameter in the necking zone which is not performed in this investigation. Instead, the true strain at fracture $\varepsilon_{f}$ can be computed from the measurement of initial $D_i$ and fracture diameter $D_f$ of the specimen according to \cite{Boehme2007}:

\begin{equation}
	\varepsilon_{f} = ln\left(\frac{A_i}{A_f}\right) = ln\left(\frac{D_i^2}{D_f^2}\right)
	\label{Glg:FractureStrain}
\end{equation}

The corresponding true stress $\sigma_f$ at fracture is then computed from the force at fracture $F_f$ and the fracture surface area $A_f$:

\begin{equation}
	\sigma_{f} = \frac{F_f}{A_f}
	\label{Glg:FractureStress}
\end{equation}

\subsection{Parameter A, B and n from Quasi-Static Tests at Room Temperature}
\label{Kap:ABn_Anpassung}

Quasi-static tensile test results at room temperature were used to derive the parameters A, B and n for the static part (first term) of the Johnson-Cook flow stress model (\ref{Glg:JohnsonCook}):

\begin{equation}
	\sigma_y^{static} = A+B \cdot \varepsilon_{pl}^n
	\label{Glg:JohnsonCook_static}
\end{equation}

The measured stresses and strains until uniform elongation $A_G$ were converted into true plastic strains and true stresses by use of equations (\ref{Glg:WahreDehnung}), (\ref{Glg:WahreSpannung}) and (\ref{Glg:WahrePlastischeDehnung}). Additionally, the stresses and strains at fracture were incorporated to the true stress- true plastic strain data. They were computed by equations (\ref{Glg:FractureStrain}) and (\ref{Glg:FractureStress}) based on measured initial $D_i$ and fracture diameter $D_f$ of the specimen and the force at fracture $F_f$, see table \ref{Tab:Fracture_stress_strain}.

\begin{table}[ht]
	\centering
	\begin{tabular}{| c | c | c | c | c | c |}
		\hline
		Specimen & $D_{i}$ & $D_{f}$ & Force at fracture $F_{f}$ & $\varepsilon_{f}$ & $\sigma_{f} $\\
		\hline
		1 & $6.00mm$ & $4.75mm$ & $18716N$ & $46.72\%$ & $1056 MPa$\\
		\hline
		2 & $5.99mm$ & $4.66mm$ & $18298N$ & $50.22\%$ & $1073 MPa$\\
		\hline
		3 & $5.99mm$ & $4.62mm$ & $18637N$ & $51.94\%$ & $1112 MPa$\\
		\hline
	\end{tabular} 
	\caption{Fracture stresses and strains from quasi-static tests}
	\label{Tab:Fracture_stress_strain}
\end{table}

The latter approach follows the proposal of \cite{Boehme2007} and shall improve predictions of the flow stress curve at higher strains towards fracture\footnote{Values at fracture were used for room temperature and quasi-static conditions only, as the influence of heating due to plastic dissipation is the lowest, in contrast to tests at higher strain rates \cite{Boehme2007}.}.

A least squares fit is used to fit the parameters A, B and n from equation (\ref{Glg:JohnsonCook_static}) to the experimental data, by minimizing the sum of the squared error of the model prediction \cite{Bronstein2005}:

\begin{equation}
	\sum_i \left[ \sigma_y^{static}(\varepsilon_{pl}^n)- \sigma_{y,i}^{measured}(\varepsilon_{pl}) \right]^2 = \sum_i \left[ A+B \cdot \varepsilon_{pl}^n- \sigma_{y,i}^{measured}(\varepsilon_{pl}) \right]^2	= min
\end{equation}

The permissible bounds for the three parameters within the least squares fit are given in table \ref{Tab:ABn_least_square_limits}.

\begin{table}[ht]
	\centering
	\begin{tabular}{| l | c | c | c | c |}
		\hline
		& A [MPa] & B [MPa] & n [-]\\% & \textbf{Comments}\\
		\hline
		Minimum & $430.9 (R_{eL}) $ & 0 & 0\\% &\\
		\hline
		Maximum & $434.5 (R_{eH})$ & 6000 & 5.999\\% &\\
		\hline
	\end{tabular} 
	\caption{Limits for the static yield stress coefficients A, B and n within the least squares fit}
	\label{Tab:ABn_least_square_limits}
\end{table}

The coefficient $A$ was limited between lower ($R_{eL}$) and upper ($R_{eH}$) yield stress from quasi-static tensile tests at room temperature and thus reflecting the initial yield stress of a virgin material\footnote{In this way, the coefficient A of the Johnson-Cook flow stress model reflects a physical meaning. However, depending on the kind of application, this requirement could be relaxed or even released.}.

\begin{table}[ht]
	\centering
	\begin{tabular}{| c | c | c | c | c |}
		\hline
		A [MPa] & B [MPa] & n [-] & $R^2$-value & Comments\\
		\hline
		 \JCA & \JCB & \JCn & 0.8262 & with fracture data\\
		 \JCAohneBruchdaten & \JCBohneBruchdaten & \JCnohneBruchdaten & - & without fracture data\\
		\hline
	\end{tabular} 
	\caption{Least squares fit of Johnson-Cook coefficients A, B \& n}
	\label{Tab:ABn_least_square_results}
\end{table}

\begin{figure}[htp]
	\center{
		\includegraphics[scale=0.8]{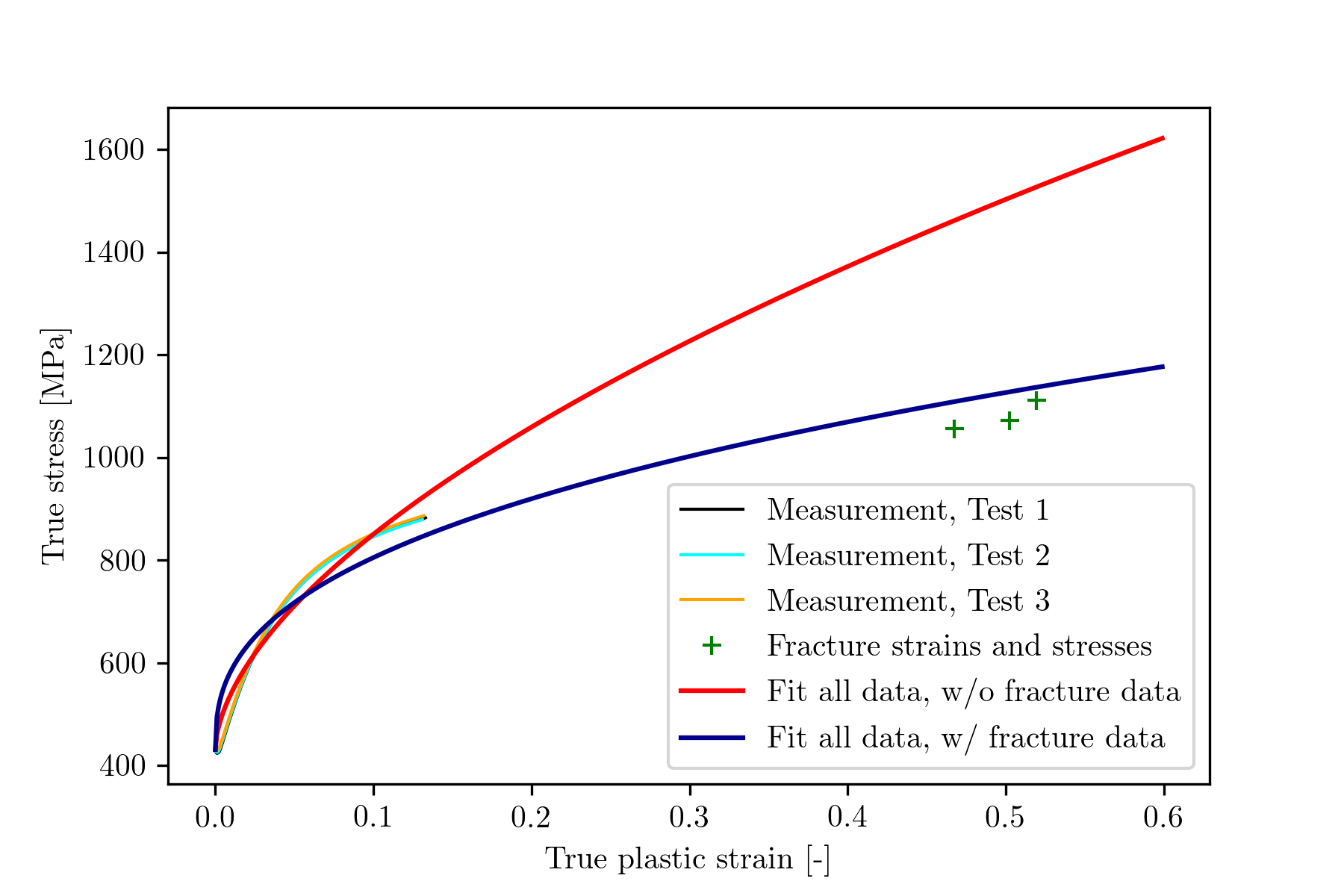}
	}
	\caption{Quasi-static tensile test results and curve fit with (blue) and without (red) consideration of fracture stress-strains}
	\label{Bild:Anpassung_ABn}
\end{figure}

The resulting parameters of the least squares fit are compiled in table \ref{Tab:ABn_least_square_results}. The first parameter set considers the fracture stress-strains, while the second set does not. Figure \ref{Bild:Anpassung_ABn} shows the measured stress-strain curves at quasi-static conditions as well as the fitted flow stress curves. The red curve (second set in table \ref{Tab:ABn_least_square_results}) represents the fit without consideration of fracture data while the blue curve considers. It can be seen that without including fracture data into the parameter fit the flow stress is predicted to be higher\footnote{at $50\%$ plastic strain, the yield stress is predicted to be 1502.9MPa instead of 1099.6MPa} at larger plastic strains and the fracture energy is significantly increased, which is expressed in the difference of the surface areas under the red and blue curve. Since the blue curve gives a better overall fit of the static yield limit part of the JC flow stress equation \eqref{Glg:JohnsonCook_static}, its parameters (first set in table \ref{Tab:ABn_least_square_results}) are used in the subsequent work.

The quality of the least squares fit is determined by the $R^2$-value\cite{Murugesan2019} which is determined by:

\begin{equation}
	R^2 = 1 - \frac{\sum\limits_i(y_i - \hat{y}_i)^2}{\sum\limits_i(y_i - \bar{y})^2}
	\label{Glg:RQuadrat_Definition}
\end{equation}

with $y_i$ the measured value, $\hat{y}_i$ the predicted value and $\bar{y}$ the mean value of the measured values. Inserting the flow stresses from the measurement $\sigma_{y,i}^{measured}(\varepsilon_{pl})$, the flow stresses from the prediction $\sigma_{y,i}^{static}(\varepsilon_{pl})$ and the measured mean value $\bar{\sigma}_{y,i}^{measured}(\varepsilon_{pl})$:

\begin{equation}
	R^2 = 1 - \frac{\sum\limits_i(\sigma_{y,i}^{measured}(\varepsilon_{pl}) - \sigma_{y,i}^{static}(\varepsilon_{pl}))^2}{\sum\limits_i(\sigma_{y,i}^{measured}(\varepsilon_{pl}) - \bar{\sigma}_{y,i}^{measured}(\varepsilon_{pl}))^2}
	\label{Glg:RQuadrat_ABn}
\end{equation}

gives $R^2=0.8262$ for the parameter set including fracture data. This is not a perfect fit but is considered to be acceptable.

\subsection{Data Preparation for Determination of Parameters C and m}

Before evaluation of parameters $C$ and $m$ all true stresses and true strains converted from equations (\ref{Glg:WahreDehnung}) and (\ref{Glg:WahreSpannung}) were smoothed because of overlayed oscillations in the measurement data, see for example figure \ref{Bild:Abq_T200_900}. Each experimental flow stress curve was first smoothed by fitting it to a polynomial, inspired by \cite{Boehme2007, Meyer2001b}. The polynomial chosen here is of the same type as the first term of the Johnson-Cook flow stress equation (\ref{Glg:JohnsonCook}):

\begin{equation}
	%\sigma_y^{static} = A+B \cdot \varepsilon_{pl}^n
	\sigma_i^{smoothed}(\varepsilon_{pl}) = a_i+b_i \cdot \varepsilon_{pl}^{c_i}
	\label{Glg:JohnsonCook_static2}
\end{equation}

where $i$ is the experiment number. The fit of the polynomial coeffcients $a_i, b_i$ and $c_i$ was performed in the range from initial yielding point until uniform strain $A_{G,i}$ using a least squares algorithm as in section \ref{Kap:ABn_Anpassung}. The permissible bounds for the coefficients $a_i, b_i$ and $c_i$ where:

\begin{table}[ht]
	\centering
	\begin{tabular}{| c | c | c | c |}
		\hline
		- & $a_i$ [MPa] & $b_i$ [MPa] & $c_i$ [-]\\
		\hline
		Minimum & $0$ & $0$ & 0\\
		\hline
		Maximum & $1200$ & $5000$ & 5.999\\
		\hline
	\end{tabular} 
	\caption{Least square fit limits for coefficients $a_i$, $b_i$ and $c_i$}
	\label{Tab:abc_least_square_limits}
\end{table}

From these polynomials flow stresses $\sigma_i^{smoothed}$ were then evaluated at a plastic strain of $\varepsilon_{pl} = 5\%$:

\begin{equation}
	\sigma_i^{smoothed}(\varepsilon_{pl}=5\%) = a_i+b_i \cdot (0.05)^{c_i}
	\label{Glg:JohnsonCook_FuenfProzentDehnung}
\end{equation}

These flow stresses are required in the following sections for the determination of the parameters C and m of the Johnson-Cook flow stress model. The coefficients $a_i, b_i$ and $c_i$, the $R^2$-value of the fit as well as the flow stress at $\varepsilon_{pl} = 5\%$ are provided with table \ref{Tab:FlowStressesAtStrains}.

\begin{table}[ht!]
	\centering
	\begin{tabular}{| c | c | c | c | c | c | c | c | c |}
		\hline
		\textbf{Test} & Temperature & Average & Uniform & $a_i$ & $b_i$ & $c_i$ & $R^2$-value & Flow Stress\\
		 i & $T_i [^\circ C]$ & Strain Rate & strain & [MPa] & [MPa]  & [-]  & $\varepsilon^{pl}\leq A_{G,i}$ & $\sigma_i^{smoothed}$ at \\
		& & \textbf{$\dot{\varepsilon}^i_{pl} [s^{-1}]$} &  $A_{G,i}[\%]$ & &  &  &  & $\varepsilon_{pl}=5\%$ [MPa]\\
		\hline
		1&$20$ & 0.001 & 13.8 & 232.1 & 1326.1 & 0.332 & 0.9855 & 723\\
		2&$20$ & 0.001 & 13.6 & 244.2 & 1321 & 0.338 & 0.9854 & 723.8\\
		3&$20$ & 0.001 & 13.7 & 249.2 & 1323.9 & 0.339 & 0.9851 & 729.2\\
		4&$20$ & 472.72 & 13.1 & 548.1 & 1313.6 & 0.631 & 0.9818 & 746.4\\
		5&$20$ & 479.09 & 14.3 & 563.7 & 1440.7 & 0.718 & 0.9719 & 731.3\\
		6&$20$ & 484.03 & 15 & 548.1 & 1134.5 & 0.616 & 0.9671 & 727.2\\
		7&$20$ & 491.11 & 13.4 & 549.6 & 1220.6 & 0.639 & 0.9836 & 729.6\\
		8&$20$ & 887.49 & 13.3 & 616.5 & 1676.7 & 0.779 & 0.9155 & 779.2\\
		9&$20$ & 894.47 & 14.5 & 606.1 & 1347 & 0.687 & 0.9478 & 778.1\\
		10&$20$ & 899.75 & 14.1 & 602.8 & 1417.1 & 0.728 & 0.9453 & 762.8\\
		11&$20$ & 907.23 & 13.3 & 588.6 & 1483.6 & 0.741 & 0.9446 & 749.7\\
		12&$20$ & 1617.94 & 13.9 & 696.5 & 2493.4 & 1.064 & 0.8076 & 799.5\\
		13&$20$ & 1642.95 & 14 & 637.8 & 1913.6 & 0.937 & 0.8551 & 753.4\\
		14&$20$ & 1678.04 & 14.7 & 700.5 & 2381.8 & 1.096 & 0.6981 & 789.7\\
		15&$20$ & 1757.72 & 9.6 & 731.8 & 2591.3 & 1.331 & 0.7851 & 779.8\\
		\hline
		16&$200$ & 454.28 & 14 & 408.6 & 969.6 & 0.52 & 0.9563 & 612.7\\
		17&$200$ & 463.12 & 14.6 & 418.6 & 998.6 & 0.512 & 0.9752 & 633.8\\
		18&$200$ & 464.06 & 13.4 & 415.9 & 1015 & 0.538 & 0.9646 & 618.5\\
		19&$200$ & 876.9 & 12.3 & 507.7 & 1391.5 & 0.731 & 0.8981 & 663.2\\
		20&$200$ & 885.76 & 13.8 & 488.5 & 1179 & 0.68 & 0.8899 & 642.3\\
		21&$200$ & 898.29 & 13.4 & 483.7 & 1180.6 & 0.695 & 0.9136 & 630.7\\
		\hline
		22&$400$ & 504.61 & 13.1 & 359.4 & 956.5 & 0.607 & 0.884 & 514.4\\
		23&$400$ & 509.95 & 12.7 & 337.5 & 860.5 & 0.544 & 0.9525 & 506.3\\
		24&$400$ & 510.56 & 12 & 333.9 & 865.8 & 0.529 & 0.9571 & 511.4\\
		25&$400$ & 938.53 & 12.2 & 422.6 & 1764.1 & 0.981 & 0.7484 & 515.8\\
		26&$400$ & 954.49 & 12.9 & 401.2 & 1082 & 0.747 & 0.8645 & 516.5\\
		27&$400$ & 954.8 & 13.8 & 381.7 & 914.1 & 0.635 & 0.8992 & 518.3\\
		\hline
		28&$600$ & 446.04 & 11.3 & 311.8 & 1188.8 & 0.522 & 0.9672 & 561\\
		29&$600$ & 468.76 & 12 & 293.7 & 1067.8 & 0.503 & 0.9748 & 530.5\\
		30&$600$ & 476.82 & 13.9 & 297.6 & 1116.6 & 0.574 & 0.9595 & 497.9\\
		31&$600$ & 747.06 & 11.7 & 548.4 & 1474.7 & 0.617 & 0.9858 & 780.9\\
		32&$600$ & 919.24 & 14.2 & 349 & 1044.4 & 0.59 & 0.9635 & 527.6\\
		33&$600$ & 924.23 & 14.4 & 330 & 1043.2 & 0.595 & 0.9672 & 505.3\\
		\hline
		34&$800$ & 444.75 & 20.4 & 235.3 & 533.6 & 0.601 & 0.9922 & 323.3\\
		35&$800$ & 448.05 & 22.6 & 238.8 & 552.7 & 0.614 & 0.9811 & 326.6\\
		36&$800$ & 471.62 & 24.3 & 191.4 & 591.3 & 0.625 & 0.9785 & 282.4\\
		37&$800$ & 893.21 & 22 & 260.4 & 692 & 0.739 & 0.9724 & 336\\
		38&$800$ & 897.57 & 22 & 248.4 & 687.8 & 0.726 & 0.976 & 326.5\\
		39&$800$ & 921.92 & 25.9 & 184.8 & 670.8 & 0.675 & 0.9829 & 273.6\\
		\hline
	\end{tabular} 
	\caption{Overview of all conducted tests. For every test the curve fit parameters $a_i, b_i$ and $c_i$, and the corresponding flow stress $\sigma_i^{smoothed}$ at $\varepsilon_{pl}=5\%$ is given.}
	\label{Tab:FlowStressesAtStrains}
\end{table}

\clearpage

\subsubsection{Parameter C}
The parameter C of the JC flow stress was fitted from flow stress measurements taken at room temperature (\Tref) and four different strain rates, corresponding to the data sets 1-15 in table \ref{Tab:FlowStressesAtStrains}. The flow stresses were evaluated at a plastic strain of $\varepsilon_{pl}=5\%$.
Each flowstress $\sigma_i^{smoothed}(\varepsilon_{pl}=5\%)$ was divided by the static yield stress $\sigma_{y}^{static} = \sigma_y(\varepsilon_{pl}=0.0, \dot \varepsilon_{pl}^{}=0.001s^{-1},T=293.15K)$ giving the yield stress ratio $r^{dyn}_{\sigma,i}$ between static and dynamic yield stress for each test:

\begin{equation}
	r^{dyn}_{\sigma,i} = \frac{\sigma_i^{smoothed}(\varepsilon_{pl}=5\%)}{\sigma_{y}^{static}(\varepsilon_{pl}=5\%)} = \left[ 1+C \cdot ln \frac{\dot\varepsilon_{pl,i}}{\dot\varepsilon^0_{pl}} \right]
	\label{Glg:JohnsonCook_ParameterC}
\end{equation}

The reference strain rate was set to the strain rate of the quasi-static tests with $\dot\varepsilon^0_{pl}=\JCepsref s^{-1}$. Finally, all yield stress ratios $r^{dyn}_{\sigma,i}$ were then used to find the parameter C by a least squares fit:

\begin{equation}
	%min_C \sum \left( r^{dyn}_{\sigma,i} - \left[ 1 + C \cdot ln \frac{\dot\varepsilon_{pl,i}}{\dot\varepsilon^0_{pl}} \right] \right)^2
	\sum_i
	\left( r^{dyn}_{\sigma,i} - \left[ 1 + C \cdot ln \frac{\dot\varepsilon_{pl,i}}{\dot\varepsilon^0_{pl}} \right] \right)^2 = min
	\label{Glg:JohnsonCook_ParameterC_FQM}
\end{equation}

\begin{figure}[h]
	\center{
		\includegraphics[scale=0.9]{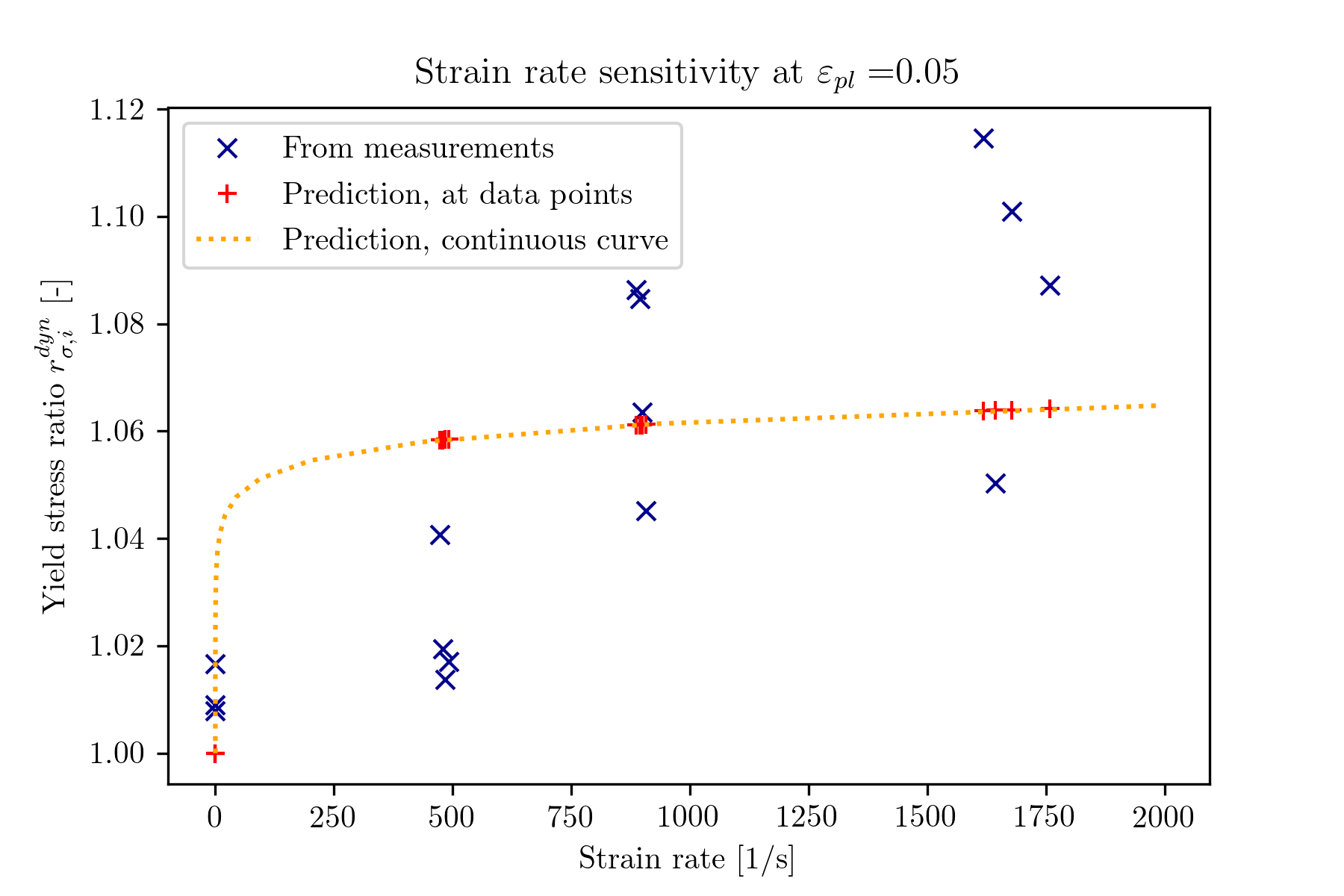}
	}
	\caption{Flow stresses at $\varepsilon_{pl}=0.05$ and various strain rates at room temperature: measured values (blue crosses), predicted at experimental strain rates (red crosses) and continuous curve (orange dotted) in the range from $\dot{\varepsilon}_{pl}=0/s..2000/s$.}
	\label{Bild:Anpassung_C_linear}
\end{figure}

\begin{figure}[h]
	\center{
		\includegraphics[scale=0.9]{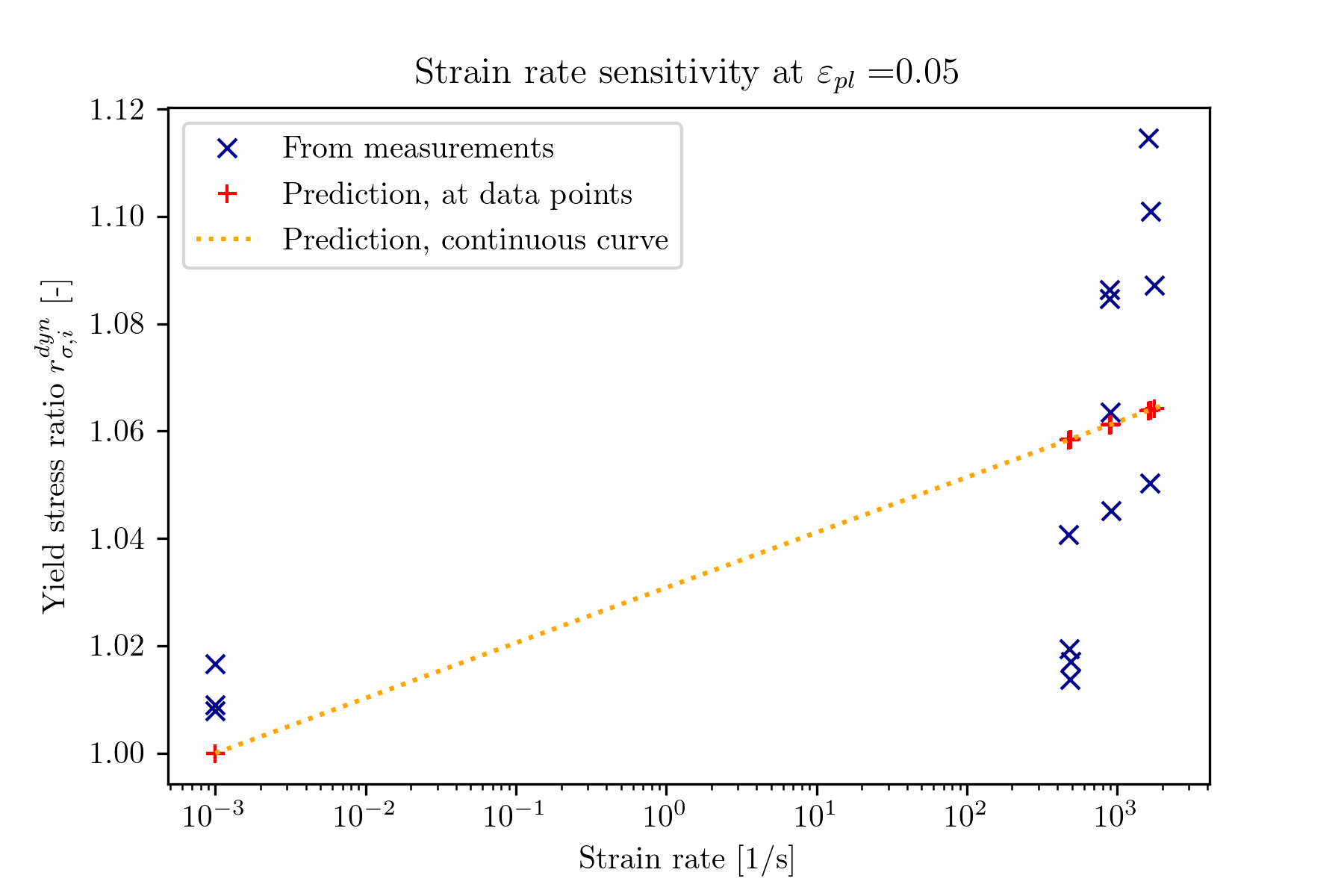}
	}
\caption{Strain rate sensitivity of the yield stress, logarithmic scale}
	\label{Bild:Anpassung_C_logEps}
\end{figure}

In figure \ref{Bild:Anpassung_C_linear} the synthetic flow stresses  from table \ref{Tab:FlowStressesAtStrains} at different strain rates are shown for a plastic strain of $\varepsilon_{pl}=5\%$ at room temperature (T=\Tref). The same results, but with a logarithmic scale of the the strain rate is shown in figure \ref{Bild:Anpassung_C_logEps}.% are shown yield stress ratios $r^{dyn}_{\sigma,i}$ are shown with logarithmic scale of the the strain rate in figure \ref{Bild:Anpassung_C_logEps}.

The fitted parameter C is given in table \ref{Tab:FinalJCParameterC} including the $R^2$-value according to equation (\ref{Glg:RQuadrat_Definition}).

\begin{table}[h]
	\centering
	\begin{tabular}{| c | c | c |}
		\hline
		\textbf{C} [-] & $\mathbf{\dot\varepsilon^0_{pl}} [s^{-1}]$ & $\mathbf{R^2} [-]$\\
		\hline
		\JCC & $\JCepsref$ & \JCCRII\\
		\hline
	\end{tabular} 
	\caption{Least squares fit of Johnson-Cook coefficient C}
	\label{Tab:FinalJCParameterC}
\end{table}

The $R^2$-value is rather low, indicating a poor fit to the experimental data as already visible in figure \ref{Bild:Anpassung_C_linear}. A similar strain rate dependency is visible also in the investigation of Thimm \cite{Thimm2019} on a C45E steel (1.1191). A straight line would give a better fit here, at least in the tested strain rate range from $0/s$ to $1700/s$, but would result in questionable predictions at higher strain rates and is therefore not followed up. It has to be noted that even with a low $R^2$-value here the overall error in the yield stress is comparably small as the strain rate sensitivity of this material is with $C=\JCC$ rather low.

\clearpage

\subsubsection{Parameter m}
\label{Kap:ParameterMAnpassung}

\begin{figure}[htp]
	\center{
		\includegraphics[scale=0.9]{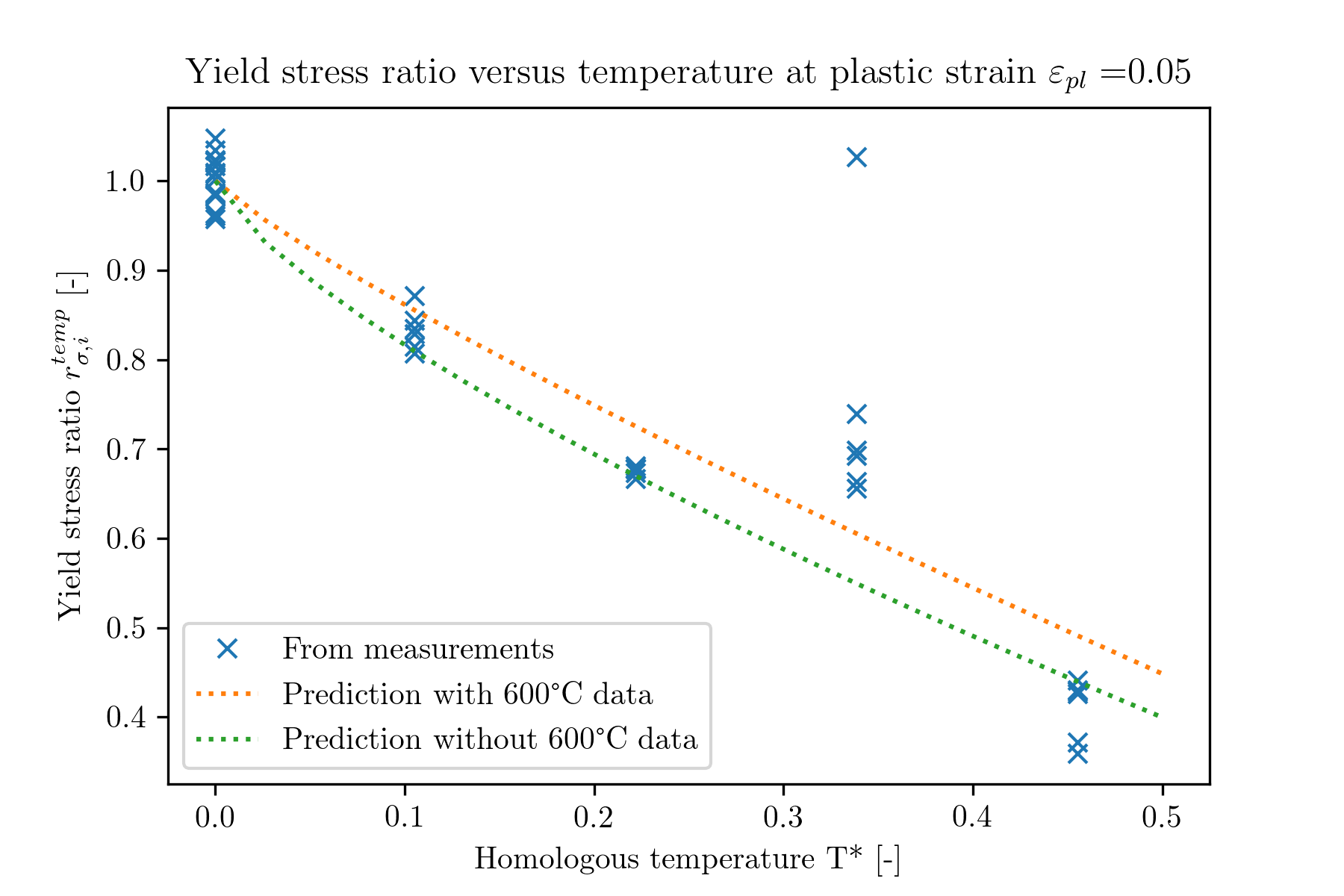}
	}
	\caption{Yield stress ratio versus homologous temperature $T^*$}
	\label{Bild:Anpassung_m}
\end{figure}

The parameter $m$ was determined similar to the strain rate sensitivity $C$ but here utilizing yield stresses $\sigma_i^{smoothed}(\varepsilon_{pl}=5\%)$ for all temperatures and strain rates, see data sets 1-39 in table \ref{Tab:FlowStressesAtStrains}. Then, the yield stress ratios $r_{\sigma,i}^{temp}$ to the first two terms of the Johnson-Cook flow stress model:

\begin{equation}
	r_{\sigma,i}^{temp} = \frac{\sigma_i^{smoothed}(\varepsilon_{pl}=5\%)}{\left( A+B \cdot \varepsilon_{pl}^n\right) \hspace{3mm} \left[ 1+C \cdot ln \frac{\dot\varepsilon_{pl}}{\dot\varepsilon^0_{pl}} \right]} = \left[1- \left(\frac{T_i-T_{ref}}{T_f-T_{ref}} \right)^m \right]	
	\label{Glg:Sigmaverhaeltnis_Parameter_m}
\end{equation}

as well as the corresponding homologous temperatures:

%\sigma_y = \left( A+B \cdot \varepsilon_{pl}^n\right) \hspace{3mm} \left[ 1+C \cdot ln \frac{\dot\varepsilon_{pl}}{\dot\varepsilon^0_{pl}} \right] \hspace{3mm} \left[1- \left(\frac{T-T_{ref}}{T_f-T_{ref}} \right)^m \right]
\begin{equation}
	T_i^* = \frac{T_i-T_{ref}}{T_f-T_{ref}}
	\label{Glg:Homologe_Temperatur}
\end{equation}

were determined, where the melting temperature is $T_f=2006K$ \cite{Smolenicki2017} and the reference temperature $T_{ref}=293.15K$ corresponds to the static tests at room temperature. Figure \ref{Bild:Anpassung_m} shows the yield stress ratios versus homologous temperature. While the general trend is a decreasing yield stress with increasing temperature, a peak exists with large scatter around $T^*=0.35$ ($T=600^\circ C$). First, the parameter m was fitted based on all data points by least squares:

\begin{equation}
	%min_m \left( r_{\sigma,i}^{temp} - \left[1- \left(T_i^* \right)^m \right] \right)^2
	\sum_i \left( r_{\sigma,i}^{temp} - \left[1- \left(T_i^* \right)^m \right] \right)^2 = min
	\label{Glg:Anpassung_m}
\end{equation}

giving the coefficient m. The fitted curve is shown in red in figure \ref{Bild:Anpassung_m}. For the temperatures $T=20^\circ C$ and $T=200^\circ C$ the fit lies in the scatter band of the measured data, while it predicts higher yield stresses at $T=400^\circ C$ and $T=800^\circ C$. At $T=600^\circ C$ the yield stress is predicted too low as the scatter of the experimental data is very large and ranges from $\approx 65\%$ to $100\%$ of the yield stress at $T=20^\circ C$. This issue is probably due to blue brittleness as discussed in \cite{Gerstgrasser2020}. The classic Johnson-Cook temperature term is not able to correctly describe this behaviour and the curve fit is worsened before and after this peak. Therefore, another curve fit (green curve) was performed, not using experimental data from $T=600^\circ C$ at all. Thus, predictions for $T=400^\circ C$ and $T=800^\circ C$ are improved, while at $T=600^\circ C$ worsened a bit. This is considered as an acceptable compromise, since the peak at $T=600^\circ C$ cannot be captured anyway.

The two fitted parameters m are given in table \ref{Tab:FinalJCParameterm} including the $R^2$-values according to equation (\ref{Glg:RQuadrat_Definition}). Beside the outlier in the range of $T^*=0.35$ ($T=600^\circ C$), the second fit can predict with acceptable accuracy the temperature characteristics and is therefore used in the following.

\begin{table}[ht]
	\centering
	\begin{tabular}{| c | c | c | c |}
		\hline
		%\textbf{m} [-] & \textbf{$R^2$}, $T=20..800^\circ C$ & \textbf{$R^2$, without $T=600^\circ C$} & Comment\\
		\textbf{m} [-] & \textbf{$R^2$} & \textbf{$R^2$} & Comment\\
		& $T=20..800^\circ C$ & \textbf{without $T=600^\circ C$} & \\
		\hline
		\JCm & \JCmRII & 0.95 & fit to all data, red curve in figure \ref{Bild:Anpassung_m}\\
		\JCmOhneSechshundertC & \JCmRIIOhneSechshundertC & 0.98 & without data at $T=600^\circ C$, green curve in figure \ref{Bild:Anpassung_m}\\
		\hline
	\end{tabular} 
	\caption{Least squares fit of Johnson-Cook coefficient m}
	\label{Tab:FinalJCParameterm}
\end{table}

\subsection{Complete Parameter Set}

All coefficients for the Johnson-Cook flow stress model are compiled in table \ref{Tab:FinalJCParameter}.

\begin{table}[ht]
	\centering
	\begin{tabular}{| c | c | c | c | c | c | c |}
		\hline
		\textbf{A} [MPa] & \textbf{B} [MPa] & \textbf{C} [-] & \textbf{m} [-] & \textbf{n} [-] & $\dot \varepsilon^{pl}_0 [s^{-1}]$ & $T_{ref}$ \\
		\hline
		%\JCA & \JCB & \JCC & \JCm & \JCn & \JCepsref & \Tref \\
		\JCA & \JCB & \JCC & \JCmOhneSechshundertC & \JCn & $\JCepsref$ & \Tref \\
		\hline
	\end{tabular} 
	\caption{Final Johnson-Cook flow stress model coefficients}
	\label{Tab:FinalJCParameter}
\end{table}

\clearpage

\section{Comparison of Measured and Predicted Flow Stresses}
\label{Kap:AnalytischerVergleich}

In this section the parameter set from table \ref{Tab:FinalJCParameter} for the Johnson-Cook flow stress model is used to compare analytical flow stress predictions versus selected stress-strain curves from the measurements. All analyzed cases are given in table \ref{Tab:SHTB_Abaqus_Simulationen}.% It is assumed that 90\% \textcolor{red}{Oder doch besser auf 0 belassen?} of the plastic work is dissipated as heat (adiabatic heating). \textcolor{red}{Evtl. Gleichung aus nächstem Kapitel schon hier bringen?}

\begin{table}[ht]
	\centering
	\begin{tabular}{| c | c | c | c | c | c |}
		\hline
		\textbf{Strain Rate $[s^{-1}]$} & $T=20^\circ C$ & $T=200^\circ C$ & $T=400^\circ C$ & $T=600^\circ C$ & $T=800^\circ C$\\
		\hline
		0.001 & $\surd$ & - & - & - & - \\
		500 & $\surd$ & - & - & - & - \\
		900 & $\surd$ & $\surd$ & $\surd$ & $\surd$ & $\surd$ \\
		1700 & $\surd$ & - & - & - & - \\
		\hline
	\end{tabular} 
	\caption{Simulated strain rates and temperatures}
	\label{Tab:SHTB_Abaqus_Simulationen}
\end{table}

\begin{figure}
	\centering
	\begin{minipage}{.5\textwidth}
		\centering
		\includegraphics[width=0.99\textwidth]{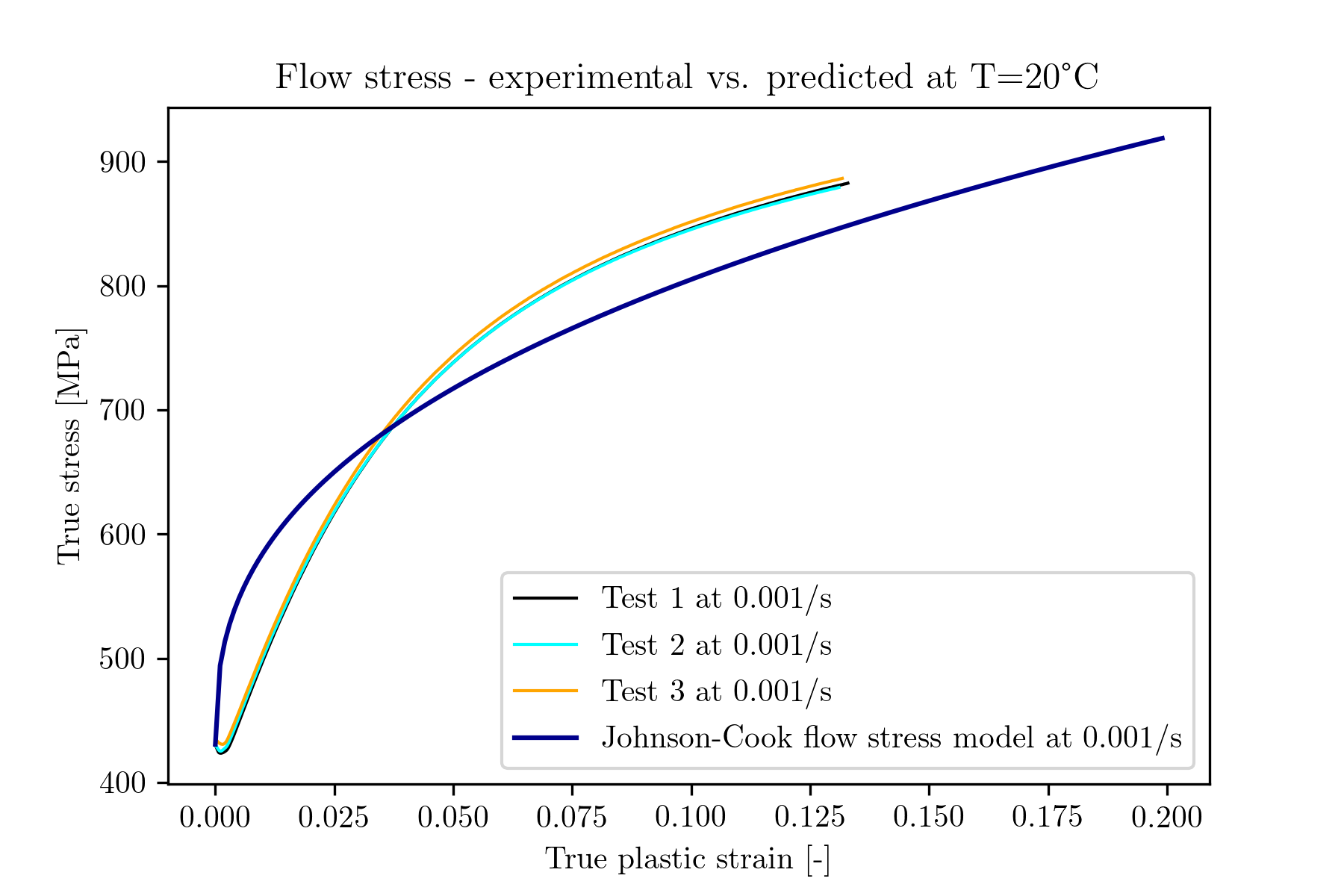}
	\end{minipage}%
	\begin{minipage}{.5\textwidth}
		\centering
		\includegraphics[width=0.99\linewidth]{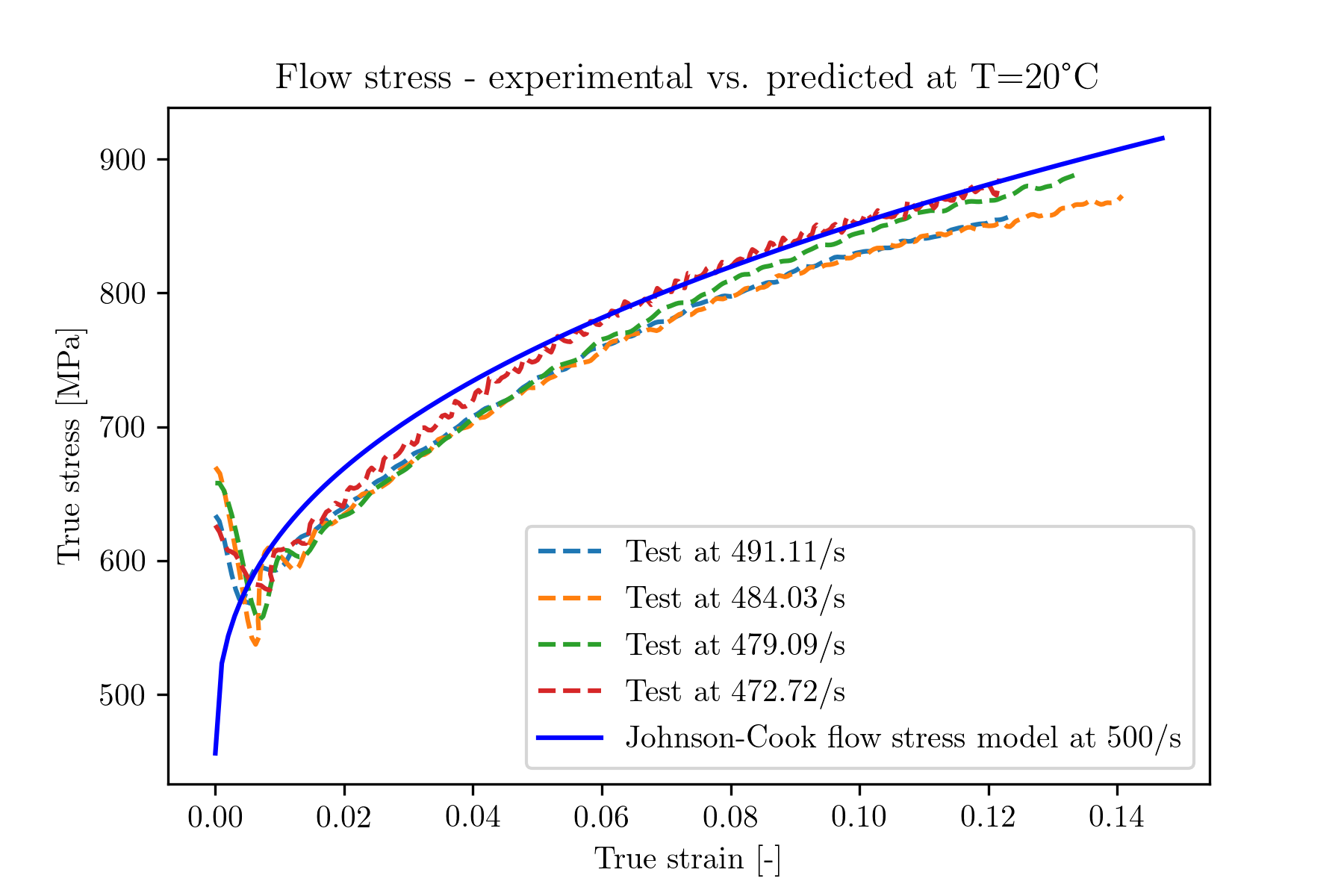}
	\end{minipage}
	\par
	\medskip
	\noindent
	\begin{minipage}[t]{.49\textwidth}
		\centering
		\captionof{figure}{Predicted and experimental flow stress curve at $T=20^\circ C, \dot{\varepsilon}_{pl}=0.001s^{-1}$}
		\label{Bild:JC20_0_001}
	\end{minipage}%
	\hfill
	\begin{minipage}[t]{.49\textwidth}
		\centering
		\captionof{figure}{Predicted and experimental flow stress curve at $T=20^\circ C, \dot{\varepsilon}_{pl}=500s^{-1}$}
		\label{Bild:JC20_500}
	\end{minipage}
\end{figure}

\begin{figure}
	\centering
	\begin{minipage}{.5\textwidth}
		\centering
		\includegraphics[width=0.99\textwidth]{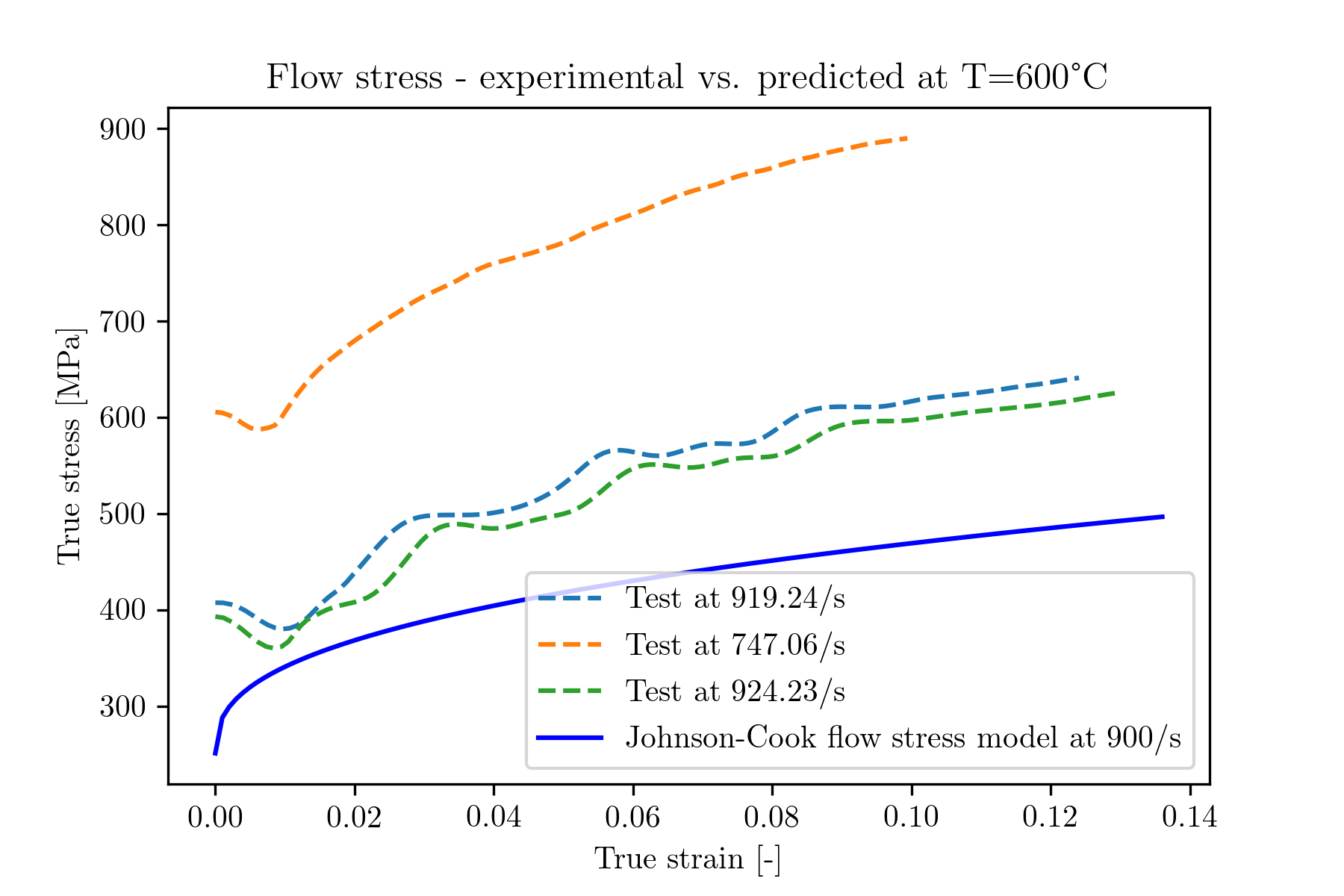}
	\end{minipage}%
	\begin{minipage}{.5\textwidth}
		\centering
		\includegraphics[width=0.99\linewidth]{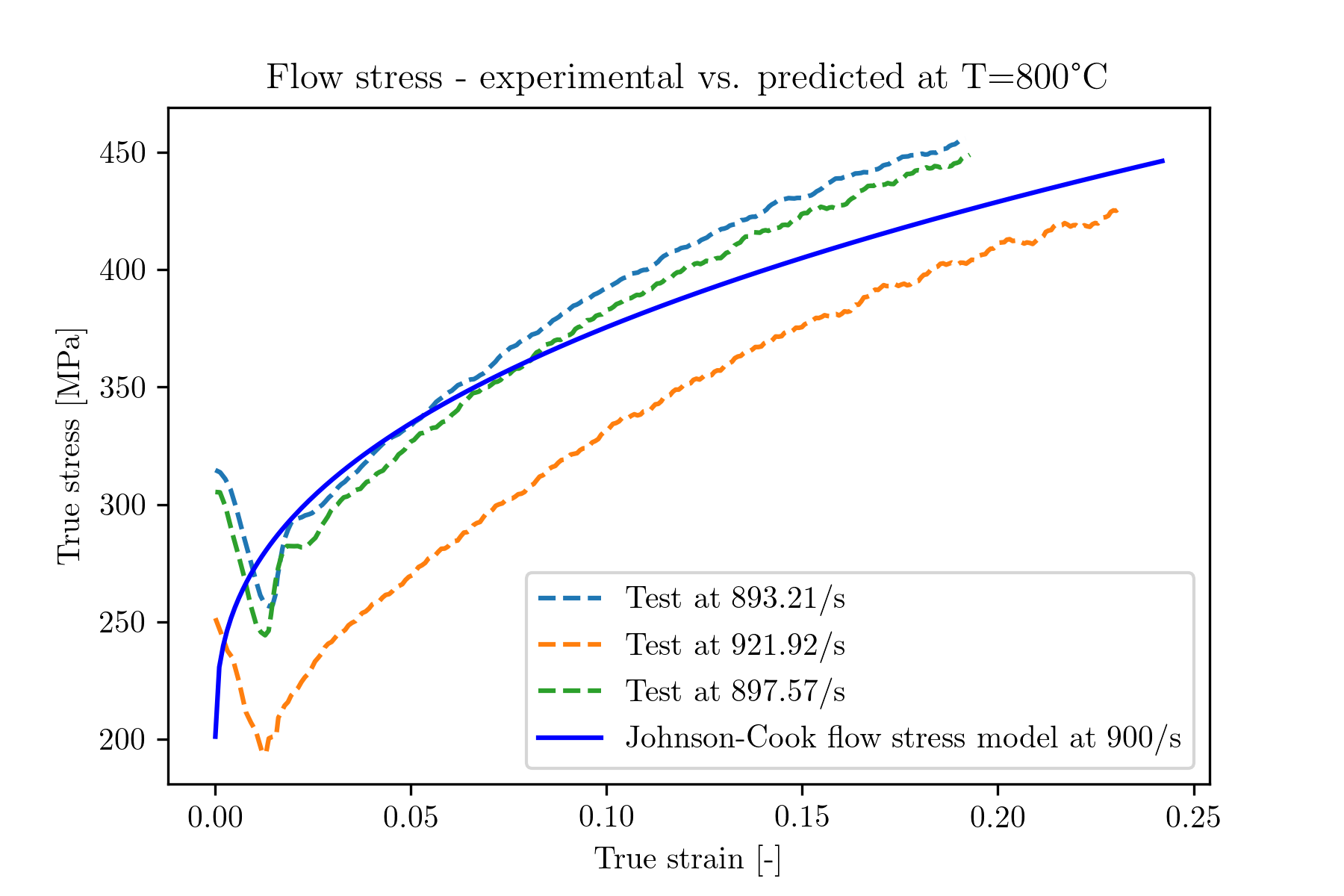}
	\end{minipage}
	\par
	\medskip
	\noindent
	\begin{minipage}[t]{.49\textwidth}
		\centering
		\captionof{figure}{Predicted and experimental flow stress curve at $T=600^\circ C, \dot{\varepsilon}_{pl}=900s^{-1}$}
		\label{Bild:JC600_900}
	\end{minipage}%
	\hfill
	\begin{minipage}[t]{.49\textwidth}
		\centering
		\captionof{figure}{Predicted and experimental flow stress curve at $T=800^\circ C, \dot{\varepsilon}_{pl}=900s^{-1}$}
		\label{Bild:JC800_900}
	\end{minipage}
\end{figure}

In this comparison the dissipation of plastic work into heat (adiabatic heating) is not considered. The graphical results are shown in figure \ref{Bild:JC20_0_001} to figure \ref{Bild:JC800_900}. In general the JC model constants from table \ref{Tab:FinalJCParameter} adequately fit to the measurements, with an exception at a temperature of $T=600^\circ C$ which was already to expect. This issue was already discussed in section \ref{Kap:ParameterMAnpassung}. Another observation is the high oscillations in the experimental data mainly at the beginning of the measurement as well as larger scatter especially at temperatures of $T=600^\circ C$ and $T=800^\circ C$

\begin{figure}
	\centering
	\begin{minipage}{.5\textwidth}
		\centering
		\includegraphics[width=0.99\textwidth]{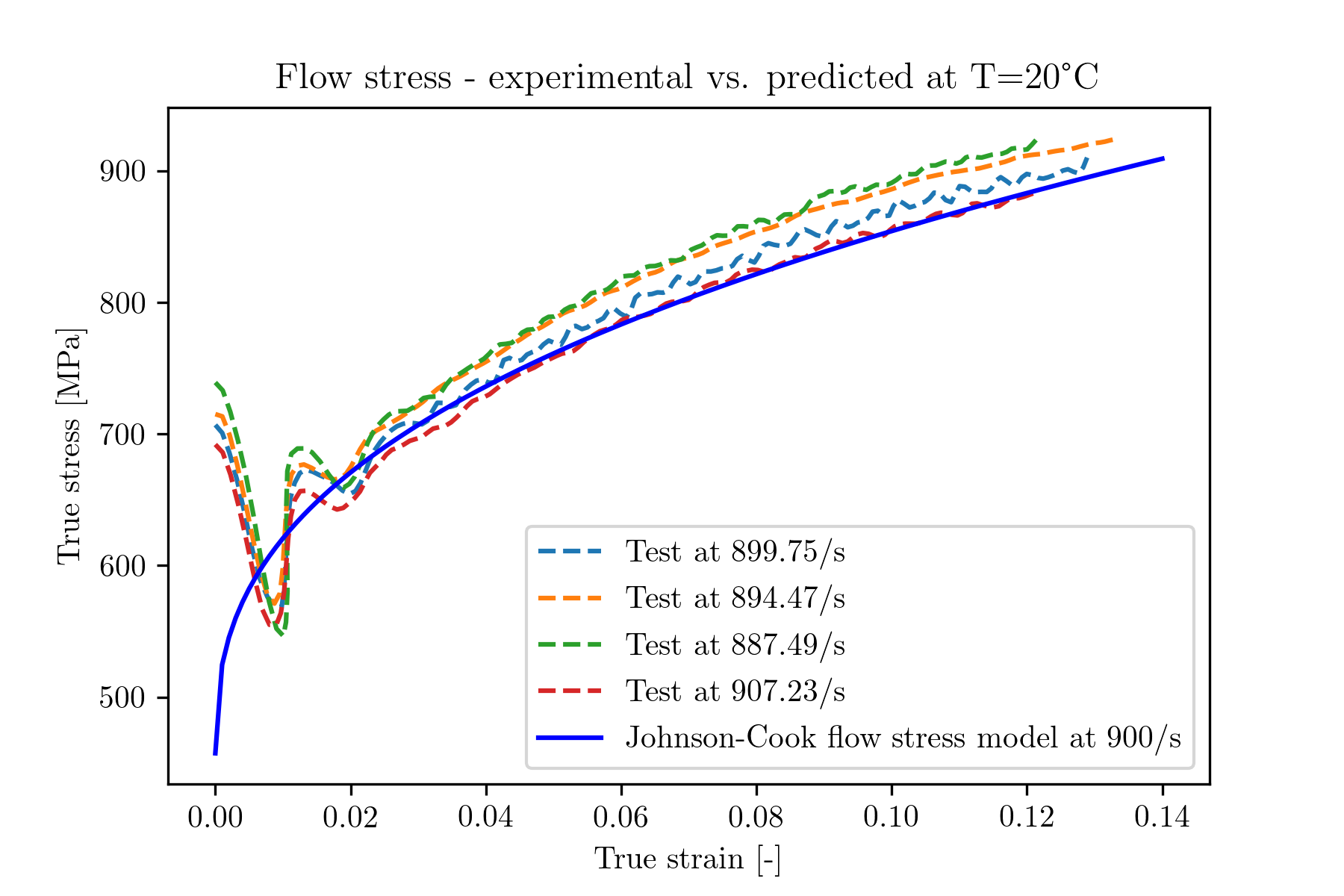}
	\end{minipage}%
	\begin{minipage}{.5\textwidth}
		\centering
		\includegraphics[width=0.99\linewidth]{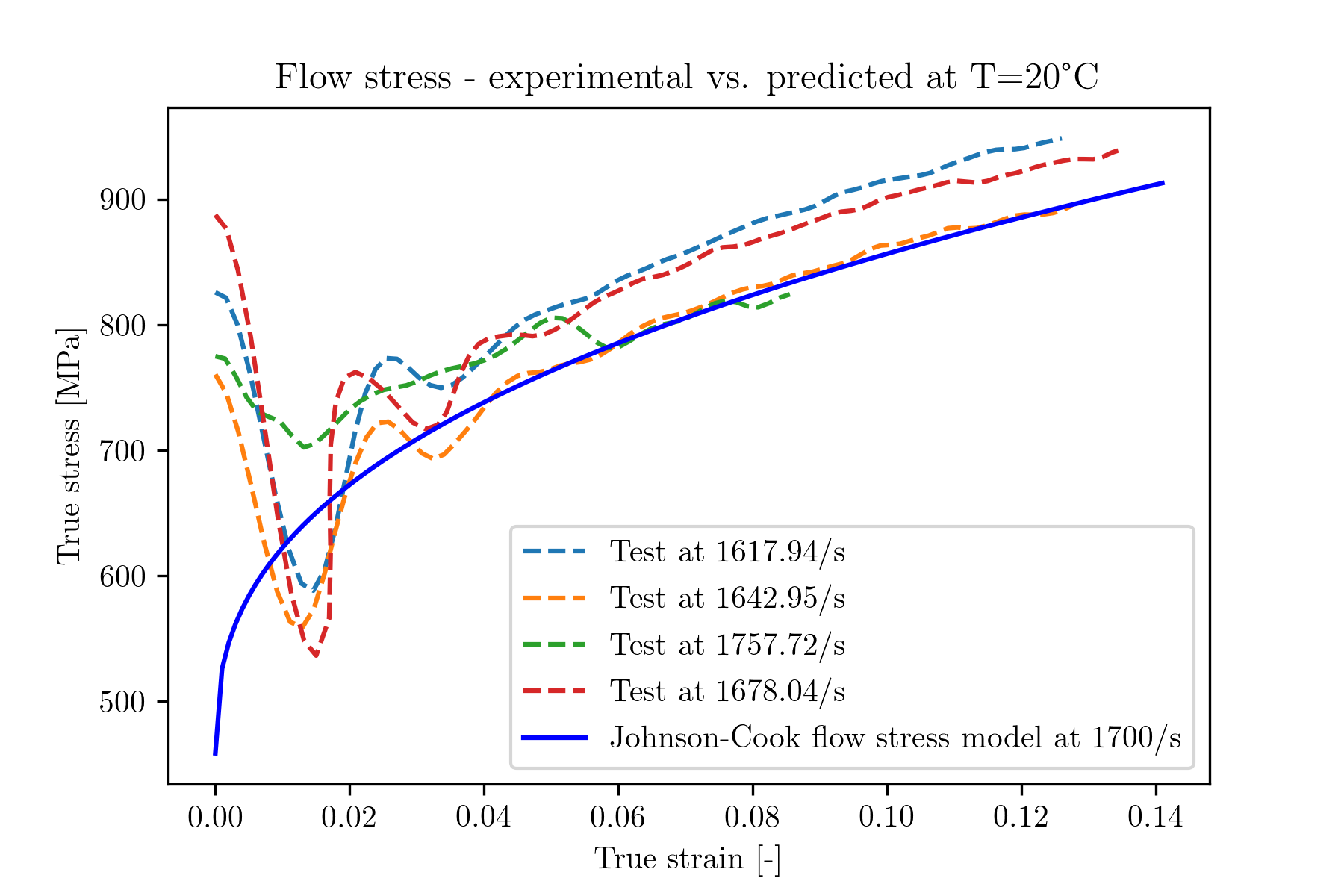}
	\end{minipage}
	\par
	\medskip
	\noindent
	\begin{minipage}[t]{.49\textwidth}
		\centering
		\captionof{figure}{Predicted and experimental flow stress curve at $T=20^\circ C, \dot{\varepsilon}_{pl}=900s^{-1}$}
		\label{Bild:JC20_900}
	\end{minipage}%
	\hfill
	\begin{minipage}[t]{.49\textwidth}
		\centering
		\captionof{figure}{Predicted and experimental flow stress curve at $T=20^\circ C, \dot{\varepsilon}_{pl}=1700s^{-1}$}
		\label{Bild:JC20_1700}
	\end{minipage}
\end{figure}

\begin{figure}
	\centering
	\begin{minipage}{.5\textwidth}
		\centering
		\includegraphics[width=0.99\textwidth]{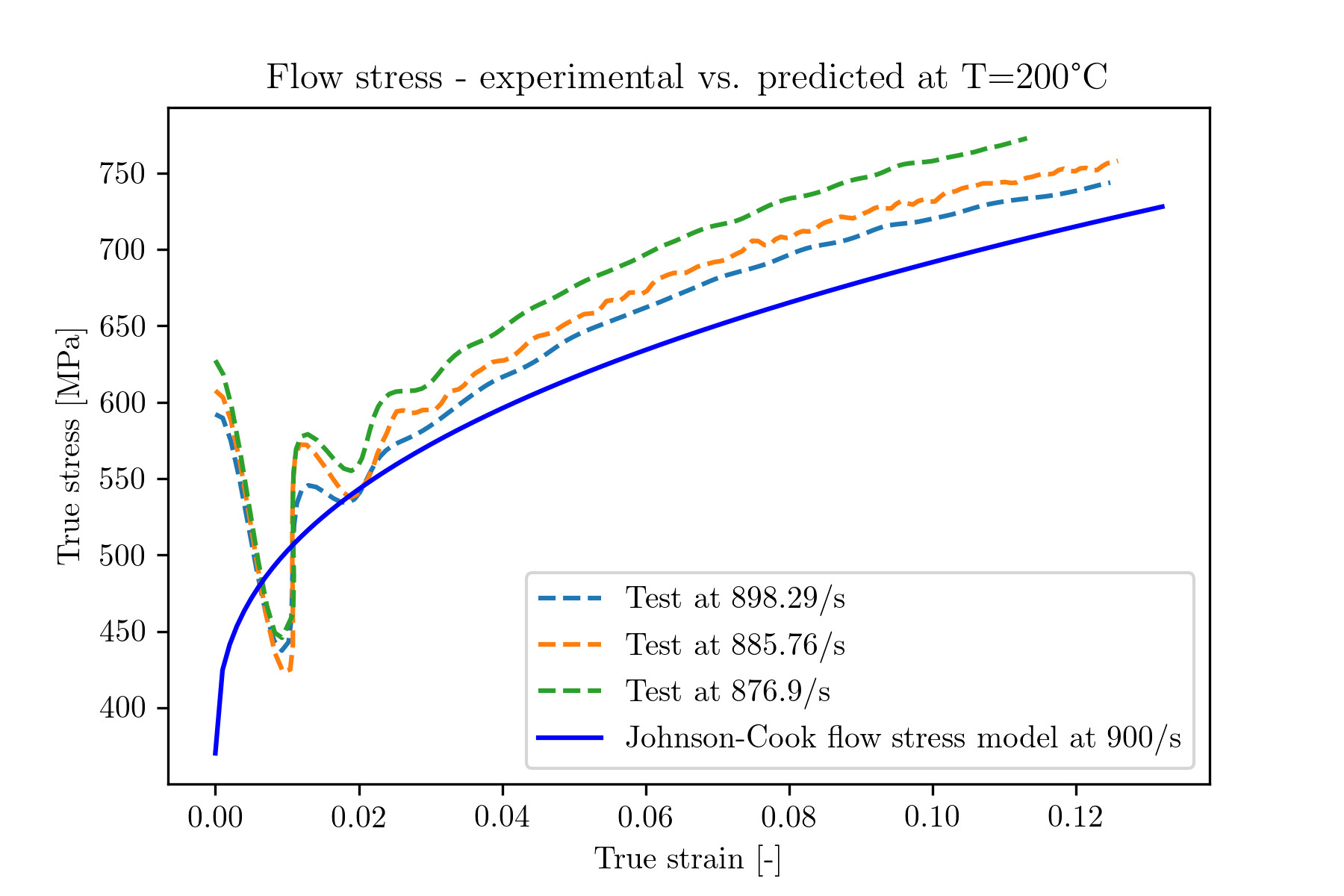}
	\end{minipage}%
	\begin{minipage}{.5\textwidth}
		\centering
		\includegraphics[width=0.99\linewidth]{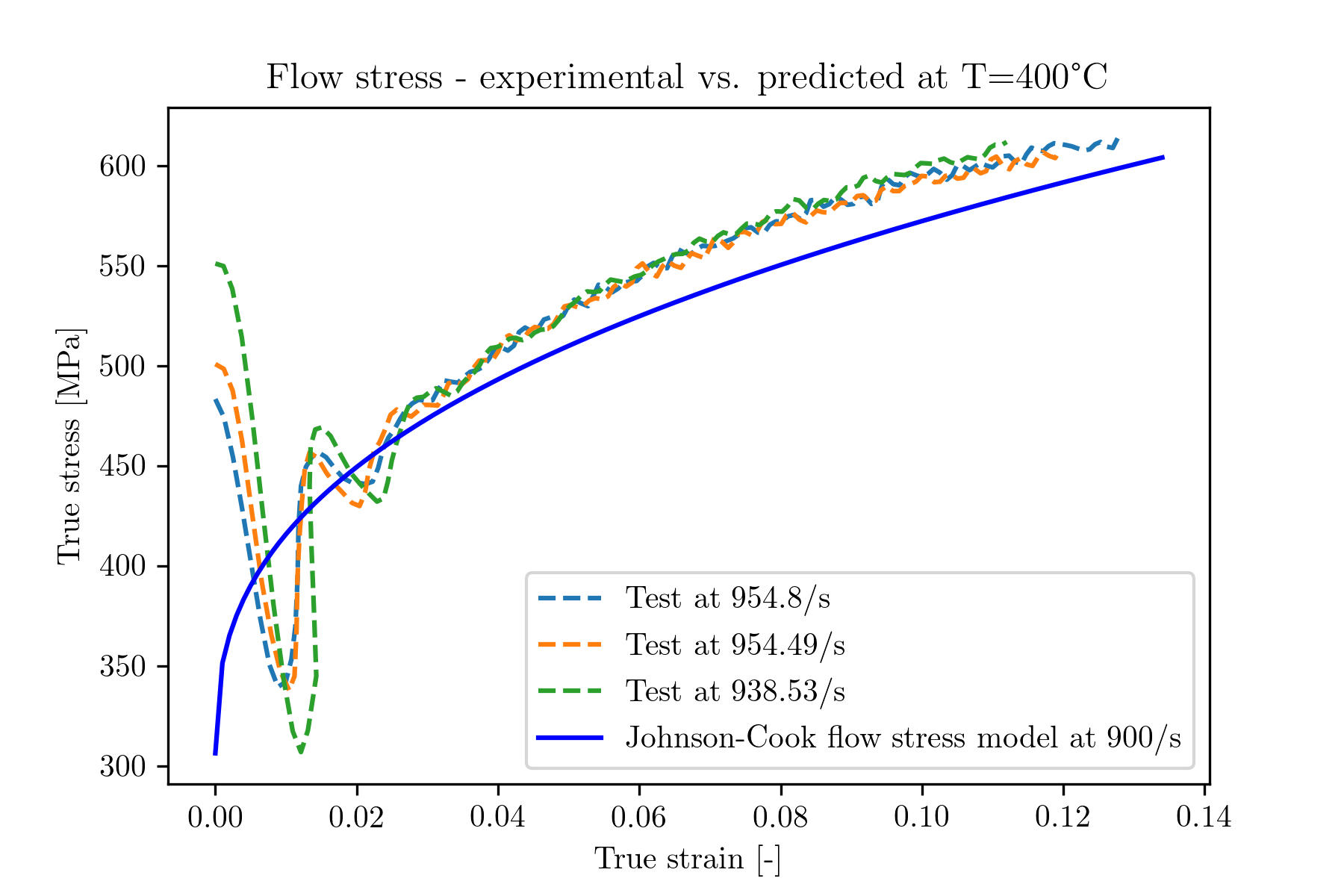}
	\end{minipage}
	\par
	\medskip
	\noindent
	\begin{minipage}[t]{.49\textwidth}
		\centering
		\captionof{figure}{Predicted and experimental flow stress curve at $T=200^\circ C, \dot{\varepsilon}_{pl}=900s^{-1}$}
		\label{Bild:JC200_900}
	\end{minipage}%
	\hfill
	\begin{minipage}[t]{.49\textwidth}
		\centering
		\captionof{figure}{Predicted and experimental flow stress curve at $T=400^\circ C, \dot{\varepsilon}_{pl}=900s^{-1}$}
		\label{Bild:JC400_900}
	\end{minipage}
\end{figure}

\clearpage

\section{Numerical Validation}
\label{Kap:Numerical_Model}

The fitted parameter set for the Johnson-Cook flow stress model was used to numerically simulate the quasi-static tensile test and the SHTB tests. The results (engineering stresses and strains) are then compared to the experimentally obtained values. The simulations were carried out with Abaqus 6-14.1 and the explicit solver.

\subsection{Geometry and Mesh}

\begin{figure}[h]
	\centering
	\begin{minipage}{.5\textwidth}
		\centering
		\includegraphics[width=0.99\textwidth]{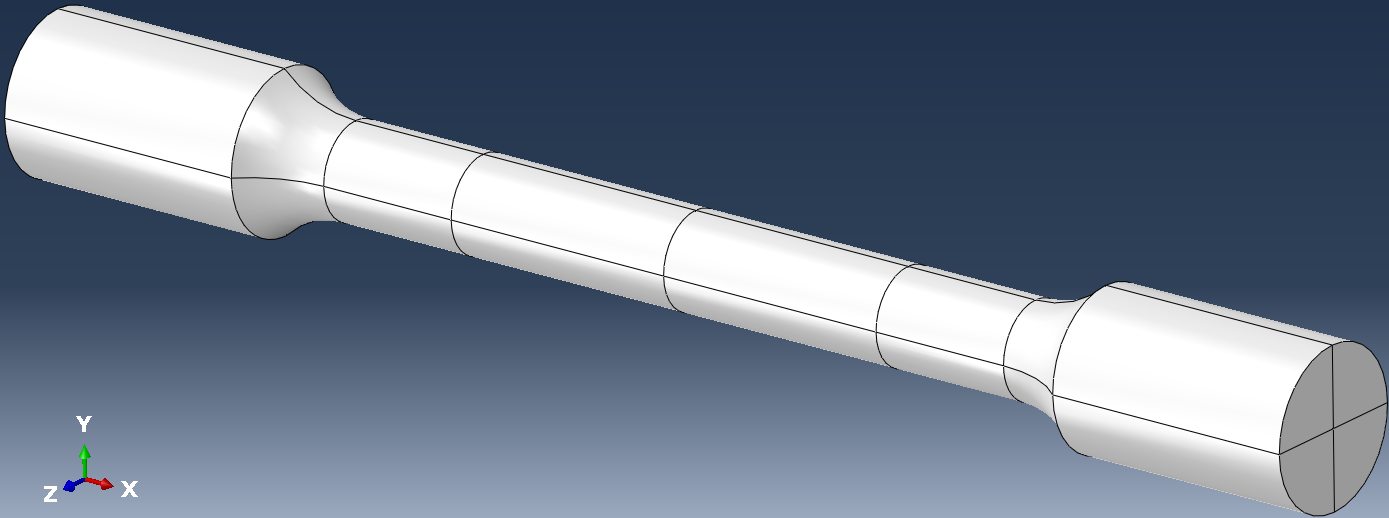}
	\end{minipage}%
	\begin{minipage}{.5\textwidth}
		\centering
		\includegraphics[width=0.99\linewidth]{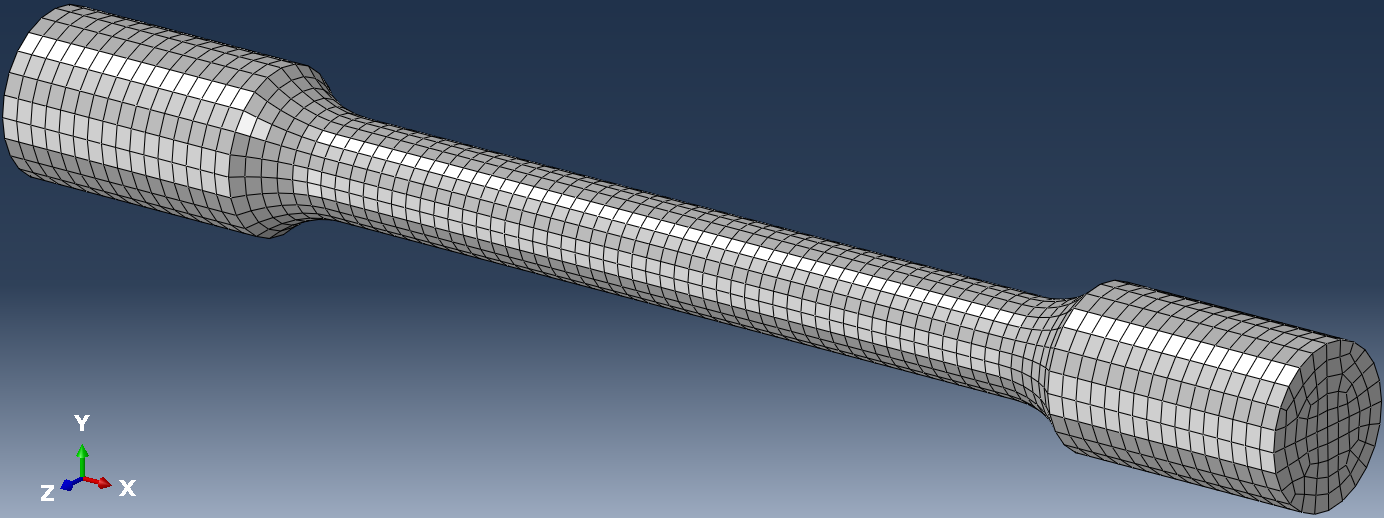}
	\end{minipage}
	\par
	\medskip
	\noindent
	\begin{minipage}[t]{.49\textwidth}
		\centering
		\captionof{figure}{Geometry of the quasi-static tensile test specimen}
		\label{Bild:AbqGeometrieQuasistatisch}
	\end{minipage}%
	\hfill
	\begin{minipage}[t]{.49\textwidth}
		\centering
		\captionof{figure}{FE-mesh of the quasi-static tensile test specimen}
		\label{Bild:AbqNetzQuasiStatisch}
	\end{minipage}
\end{figure}

\begin{figure}[h]
	\centering
	\begin{minipage}{.5\textwidth}
		\centering
		\includegraphics[width=0.99\linewidth]{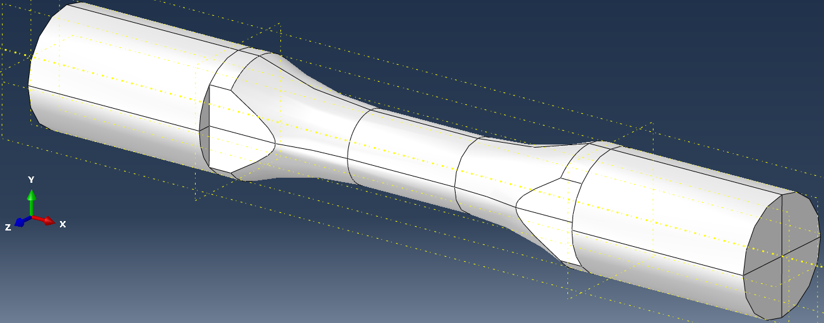}
	\end{minipage}%
	\begin{minipage}{.5\textwidth}
		\centering
		\includegraphics[width=0.99\textwidth]{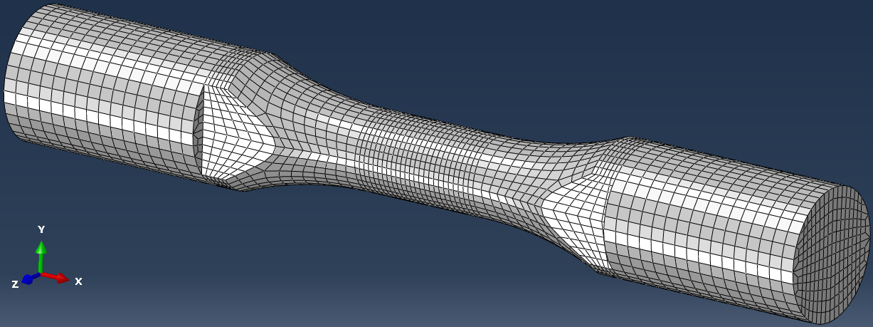}
	\end{minipage}
	\par
	\medskip
	\noindent
	\begin{minipage}[t]{.49\textwidth}
		\centering
		\captionof{figure}{Geometry of the SHTB test specimen}
		\label{Bild:AbqGeometrieSHTB}
	\end{minipage}
	\hfill
	\begin{minipage}[t]{.49\textwidth}
		\centering
		\captionof{figure}{FE-mesh of the SHTB test specimen}
		\label{Bild:AbqNetzSHTB}
	\end{minipage}
\end{figure}

%The Abaqus 6-14.1/Explicit solver was used for the numerical simulation of the tensile tests.
The geometries were built in Abaqus/CAE according to specimen drawings from figure \ref{Bild:Quasistatisch} and figure \ref{Bild:STHB_Zeichnung_SUPSI} and are shown in figure \ref{Bild:AbqGeometrieQuasistatisch} and \ref{Bild:AbqGeometrieSHTB}. The geometries were meshed with elements of the type C3D8R. The quasi-static tensile test specimen consists of 7084 elements with 8370 nodes, a picture is provided with figure \ref{Bild:AbqNetzQuasiStatisch}. The SHTB-test specimen consists of 14032 elements with 16154 nodes, a picture is provided with figure \ref{Bild:AbqNetzSHTB}.

\subsubsection{Material Parameters}

The test specimen material is 50SiB8. As described in the introduction a flow stress model according to Johnson and Cook \cite{JohnsonCook1983} is used. All material parameters used throughout the analysis are provided with tables \ref{Tab:FinalJCParameter} and \ref{Tab:Materialparameter_Workpiece}.

\begin{table}[ht]
	\centering
	\begin{tabular}{| l | c | c | c | c | c ||}
		\hline
		\textbf{Parameter} & \textbf{Symbol} & \textbf{Value} & \textbf{Unit} & \textbf{Source} & \textbf{Comments}\\
		\hline
		Density & $\rho$ & 7850 & $\frac{kg}{m^3}$ & \cite{Smolenicki2017} &\\
		Modulus of elasticity & $E$ & 214 & $GPa$ & \cite{Smolenicki2017} & value rounded\\
		Shear modulus & $G$ & 80 & $GPa$ & \cite{Smolenicki2017} & \\
		Poisson ratio & $\nu$ & 0.334875 & $-$ & - & deduced from $E$ and $G$\\
		Specific heat capacity & $c_p$ & 466 & $\frac{J}{kgK}$ & \cite{Smolenicki2017} &\\
		Melting temperature & $T_f$ & 2006 & $K$ & \cite{Smolenicki2017} & for JC\\ % Wird weder von Roelofs, Akbari oder Smolenicki erwähnt! Diskussion mit Marcel GG per Whatsapp -> Dissertation D. Smolenicki referenzieren, Do, 23.04.2020
		\hline
	\end{tabular} 
	\caption{Material properties of 50SiB8 used in the analysis}
	\label{Tab:Materialparameter_Workpiece}
\end{table}

Plastic dissipation into thermal energy (adiabatic heating) was considered with a Taylor-Quinney coefficient of $\eta_{TQ}=0.90$:

\begin{equation}
	\Delta T = \frac{\eta_{TQ} \cdot	\sigma_y}{\rho \cdot c_p} \Delta \varepsilon_{pl}
\end{equation}

The temperature dependencies of the elastic modulus, the density and the Poisson's ratio as well as heat conduction were not considered in the present work.

\subsection{Boundary Conditions}

\begin{figure}[h]
	\centering
	\begin{minipage}{.5\textwidth}
		\centering
		\includegraphics[width=0.99\linewidth]{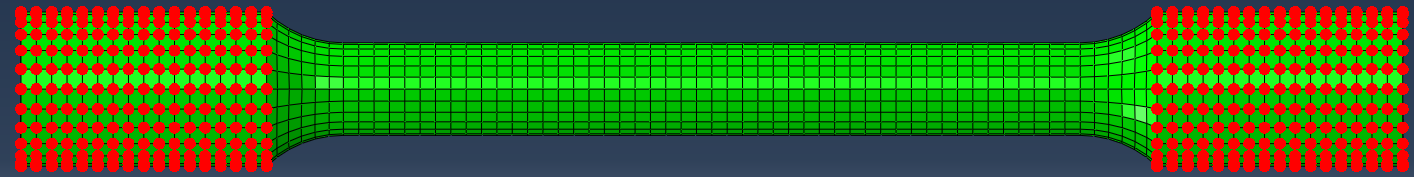}
	\end{minipage}%
	\begin{minipage}{.5\textwidth}
		\centering
		\includegraphics[width=0.8\textwidth]{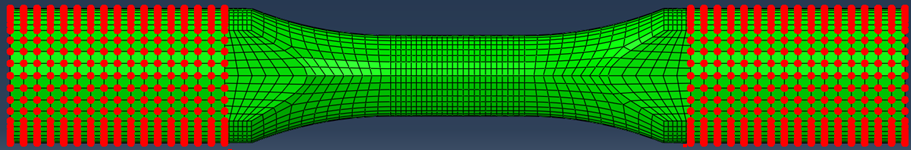}
	\end{minipage}
	\par
	\medskip
	\noindent
	\begin{minipage}[t]{.49\textwidth}
		\centering
		\captionof{figure}{Left and right boundary condition application regions marked with red dots}
		\label{Bild:GeklemmteLaengeQuasistatisch}
	\end{minipage}
	\hfill
	\begin{minipage}[t]{.49\textwidth}
		\centering
		\captionof{figure}{Left and right boundary condition application region marked with red dots}
		\label{Bild:GeklemmteLaengeSHTB}
	\end{minipage}
\end{figure}

Time dependent displacements are prescribed on the left and right side of the test specimen to reflect the strain rate of the testing. The boundary condition application regions are shown in figure \ref{Bild:GeklemmteLaengeQuasistatisch} and \ref{Bild:GeklemmteLaengeSHTB}. At room temperature four different strain rates were simulated.% and their corresponding velocities at left and right fixture were computed based on the clamped length between left and right fixture, see figure \ref{Bild:GeklemmteLaengeQuasistatisch} \ref{Bild:GeklemmteLaengeSHTB}.

Another five simulations were performed at a strain rate of $\dot{\varepsilon}_{pl}=900/s$ and varying temperatures. The test specimens' temperatures were initialized according to the temperatures given in table \ref{Tab:SHTB_Abaqus_Simulationen}. All simulations conducted are summarized in table \ref{Tab:SHTB_Abaqus_Simulationen}.

\subsection{Results}
\label{Kap:Results}

%A selection of experiments according to table \ref{Tab:SHTB_Abaqus_Simulationen} 
The resulting engineering stress-strain relations from the numerical simulations were compared to the experimental results. The engineering stress was computed from the tensile force $F$ related to the initial specimen diameter $d_0$:

\begin{equation}
	\sigma_{eng} = \frac{F}{A_0} = \frac{F}{\pi \cdot d_0^2/4}
	\label{Glg:TechnischeSpannungAbaqus}
\end{equation}

The engineering strain was computed from the current length $l$ related to the initial length $l_0$:

\begin{equation}
	\varepsilon_{eng} = \frac{\Delta l}{l_0} = \frac{l - l_0}{l_0} = \frac{l}{l_0} - 1
	\label{Glg:TechnischeDehnungAbaqus}
\end{equation}

The initial length used for the quasi-static test specimen is the gauge length of $l_0=48mm$ and for the SHTB test specimen of $l_0=5mm$.

\begin{figure}
	\centering
	\begin{minipage}{.5\textwidth}
		\centering
		\includegraphics[width=0.99\textwidth]{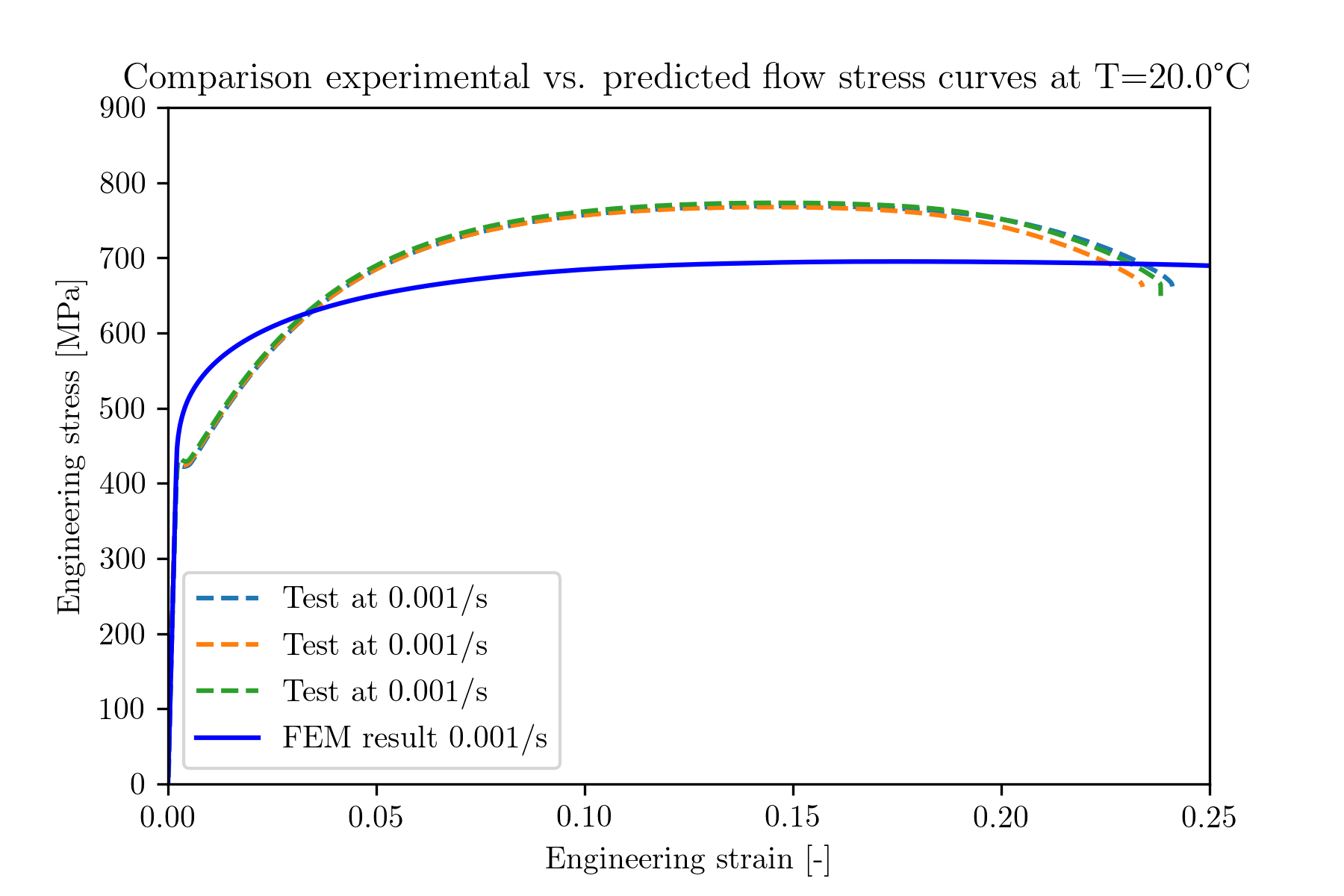}
	\end{minipage}%
	\begin{minipage}{.5\textwidth}
		\centering
		\includegraphics[width=0.99\linewidth]{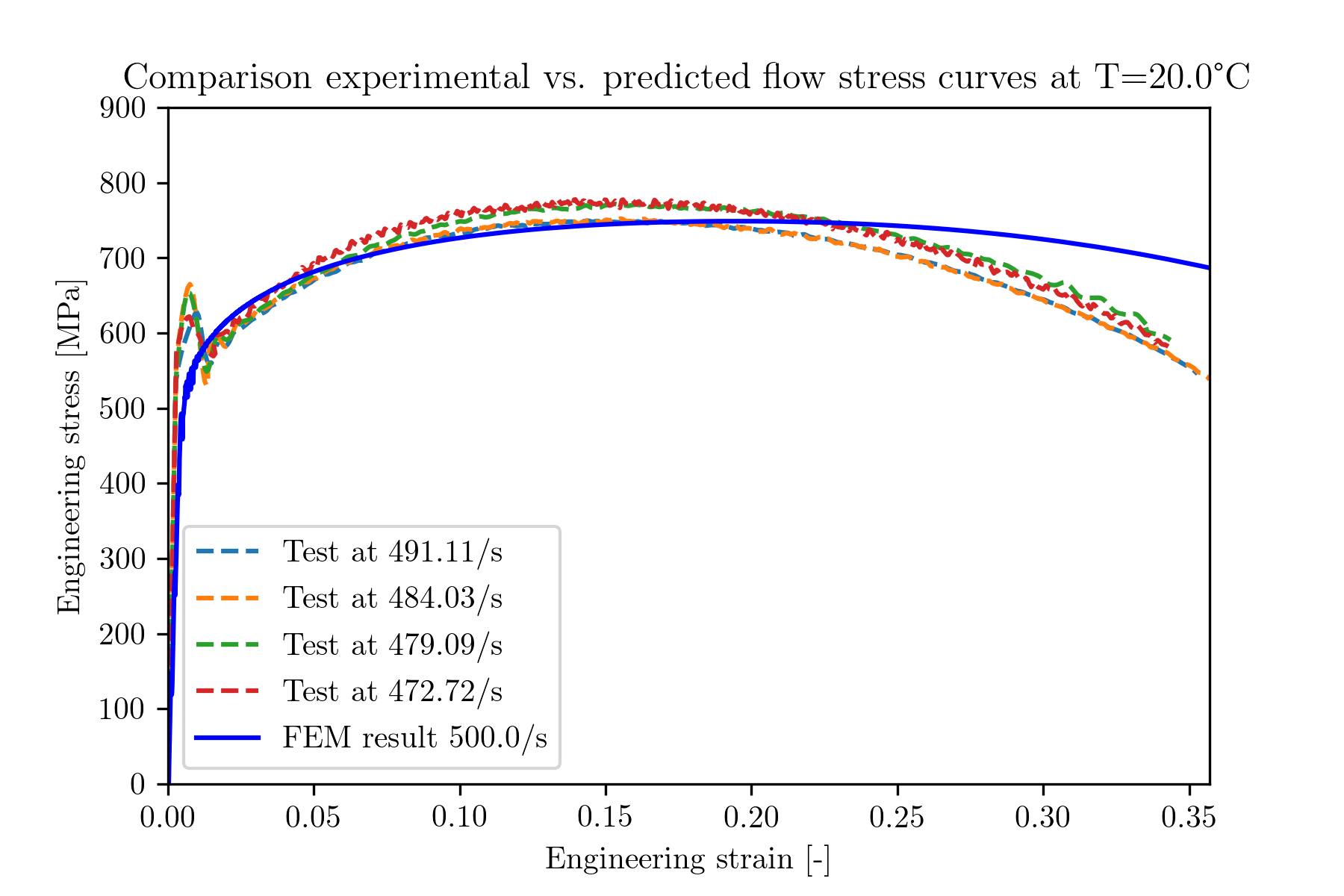}
	\end{minipage}
	\par
	\medskip
	\noindent
	\begin{minipage}[t]{.49\textwidth}
		\centering
		\captionof{figure}{Numerical and experimental flow stress curve at $T=20^\circ C, \dot{\varepsilon}_{pl}=0.001s^{-1}$}
		\label{Bild:Abq_T20_0_001}
	\end{minipage}%
	\hfill
	\begin{minipage}[t]{.49\textwidth}
		\centering
		\captionof{figure}{Numerical and experimental flow stress curve at $T=20^\circ C, \dot{\varepsilon}_{pl}=500s^{-1}$}
		\label{Bild:Abq_T20_500}
	\end{minipage}
\end{figure}

\begin{figure}
	\centering
	\begin{minipage}{.5\textwidth}
		\centering
		\includegraphics[width=0.99\textwidth]{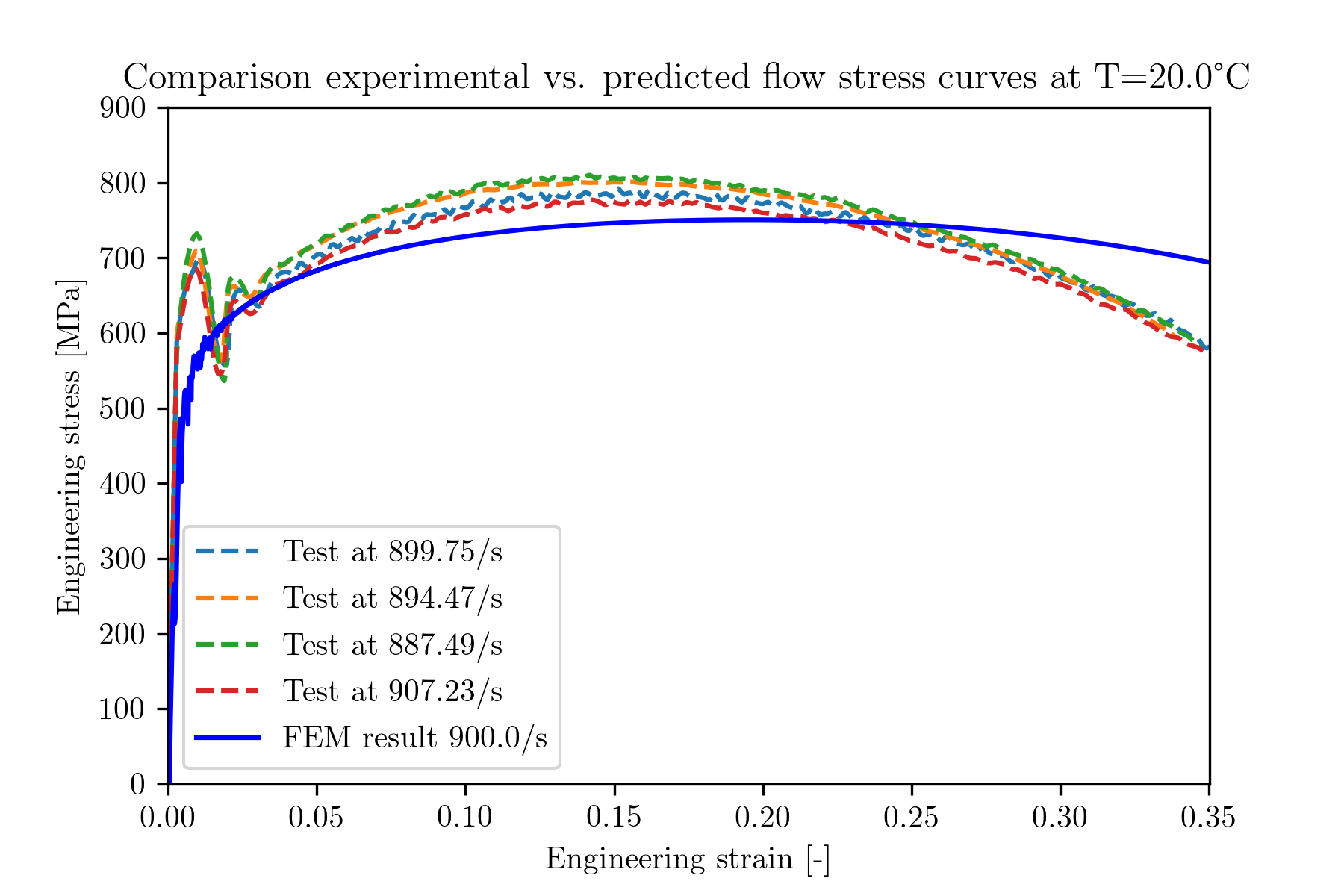}
	\end{minipage}%
	\begin{minipage}{.5\textwidth}
		\centering
		\includegraphics[width=0.99\linewidth]{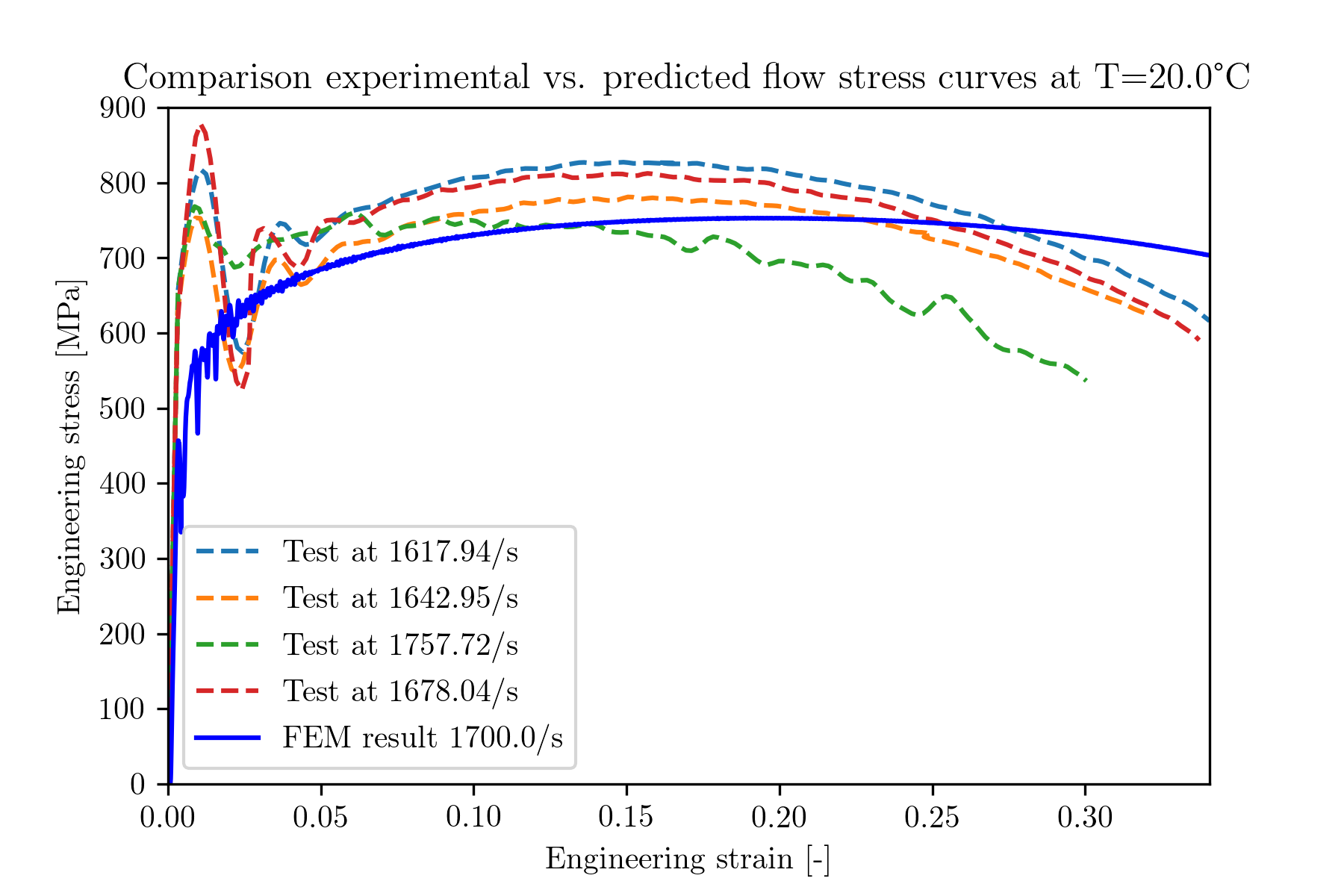}
	\end{minipage}
	\par
	\medskip
	\noindent
	\begin{minipage}[t]{.49\textwidth}
		\centering
		\captionof{figure}{Numerical and experimental flow stress curve at $T=20^\circ C, \dot{\varepsilon}_{pl}=900s^{-1}$}
		\label{Bild:Abq_T20_900}
	\end{minipage}%
	\hfill
	\begin{minipage}[t]{.49\textwidth}
		\centering
		\captionof{figure}{Numerical and experimental flow stress curve at $T=20^\circ C, \dot{\varepsilon}_{pl}=1700s^{-1}$}
		\label{Bild:Abq_T20_1700}
	\end{minipage}
\end{figure}

\begin{figure}
	\centering
	\begin{minipage}{.5\textwidth}
		\centering
		\includegraphics[width=0.99\textwidth]{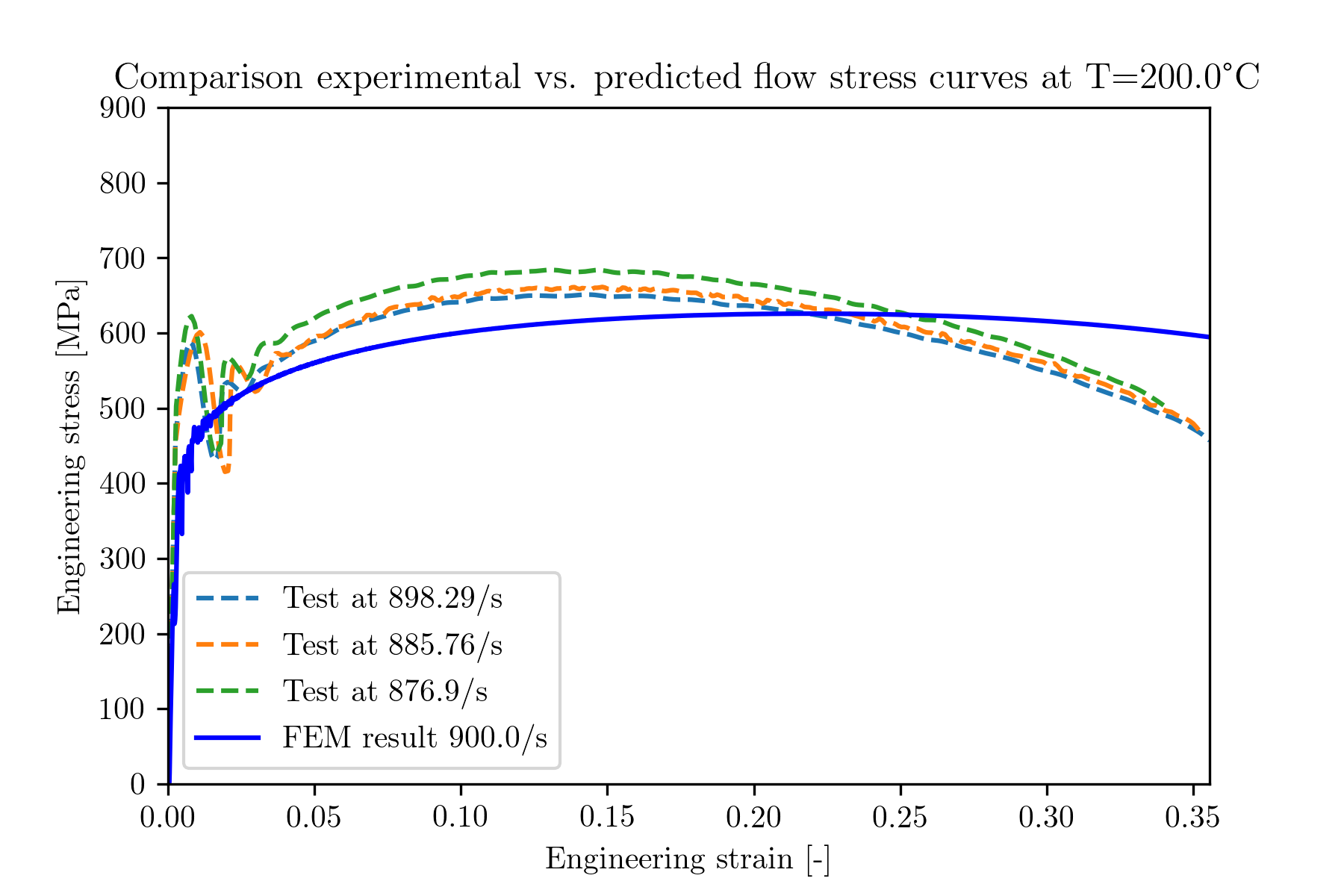}
	\end{minipage}%
	\begin{minipage}{.5\textwidth}
		\centering
		\includegraphics[width=0.99\linewidth]{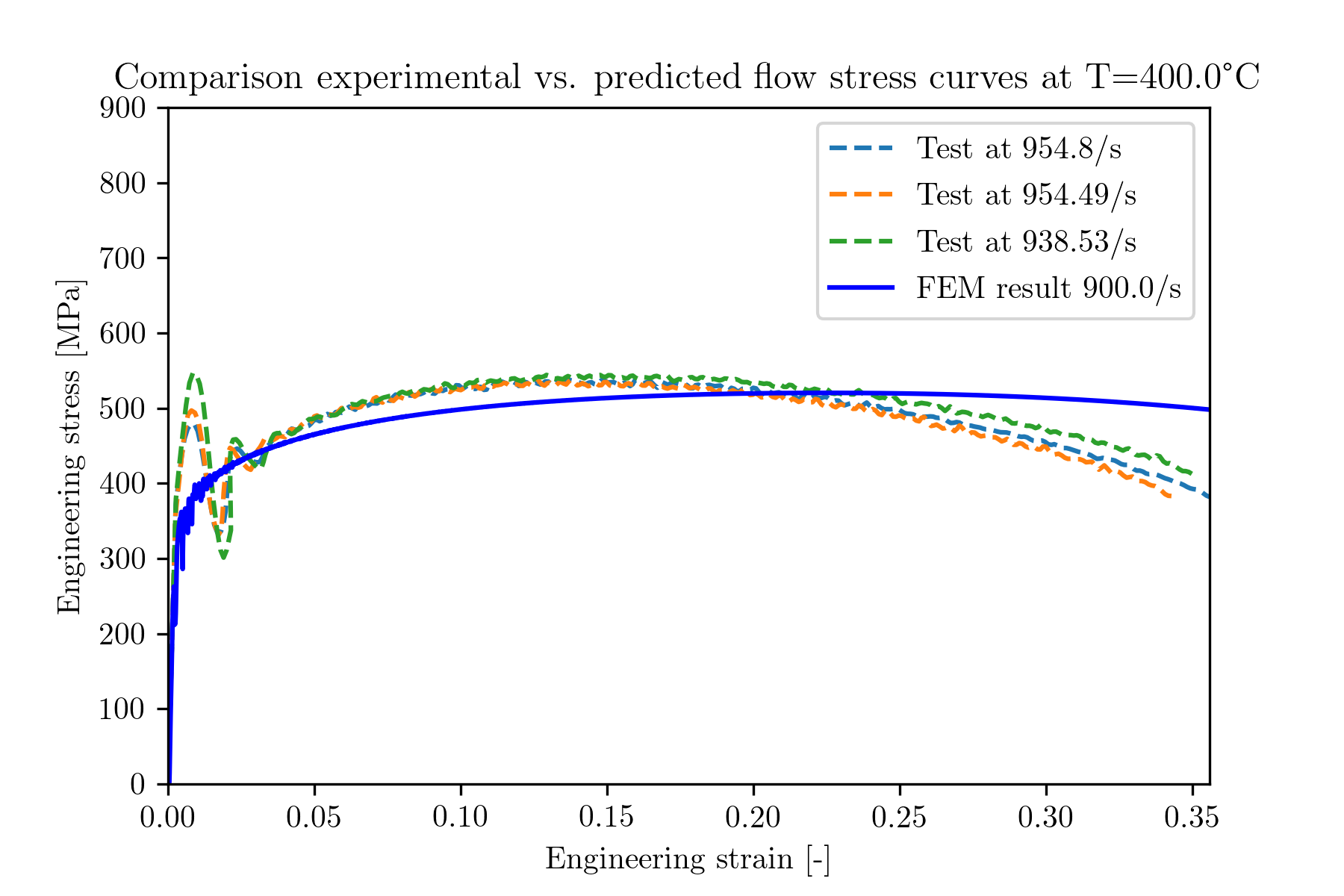}
	\end{minipage}
	\par
	\medskip
	\noindent
	\begin{minipage}[t]{.49\textwidth}
		\centering
		\captionof{figure}{Numerical and experimental flow stress curve at $T=200^\circ C, \dot{\varepsilon}_{pl}=900s^{-1}$}
		\label{Bild:Abq_T200_900}
	\end{minipage}%
	\hfill
	\begin{minipage}[t]{.49\textwidth}
		\centering
		\captionof{figure}{Numerical and experimental flow stress curve at $T=400^\circ C, \dot{\varepsilon}_{pl}=900s^{-1}$}
		\label{Bild:Abq_T400_900}
	\end{minipage}
\end{figure}

\begin{figure}
	\centering
	\begin{minipage}{.5\textwidth}
		\centering
		\includegraphics[width=0.99\textwidth]{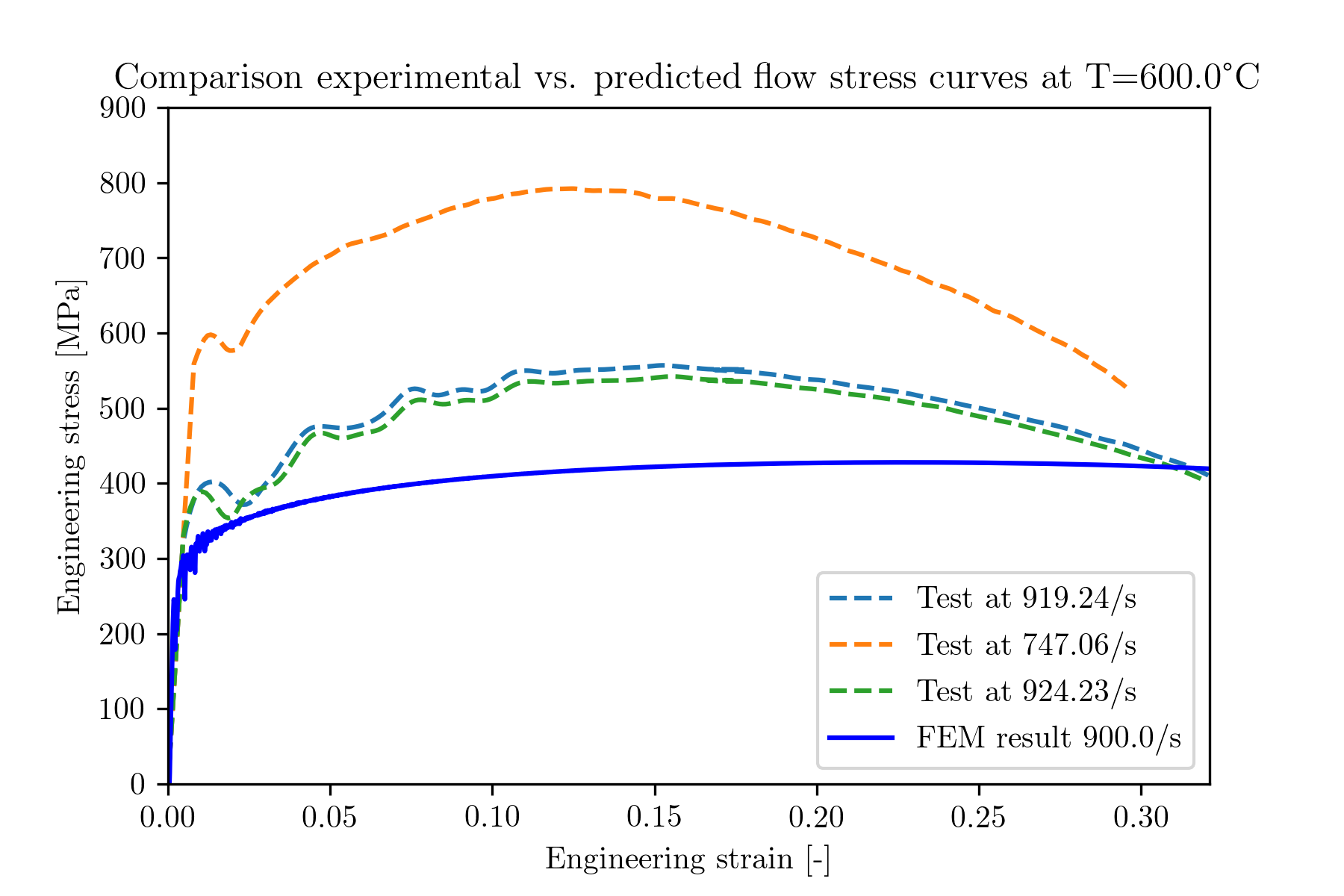}
	\end{minipage}%
	\begin{minipage}{.5\textwidth}
		\centering
		\includegraphics[width=0.99\linewidth]{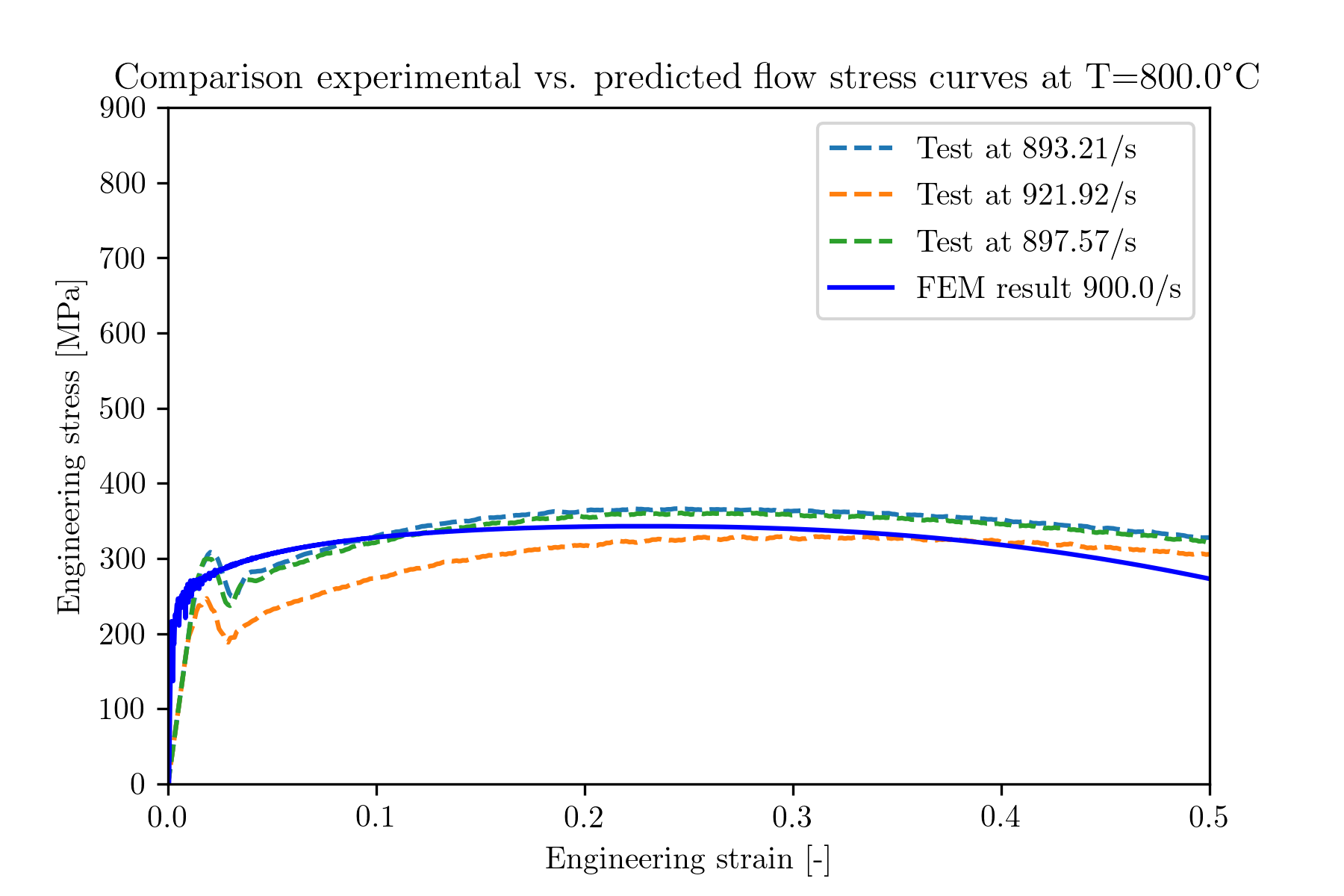}
	\end{minipage}
	\par
	\medskip
	\noindent
	\begin{minipage}[t]{.49\textwidth}
		\centering
		\captionof{figure}{Numerical and experimental flow stress curve at $T=600^\circ C, \dot{\varepsilon}_{pl}=900s^{-1}$}
		\label{Bild:Abq_T600_900}
	\end{minipage}%
	\hfill
	\begin{minipage}[t]{.49\textwidth}
		\centering
		\captionof{figure}{Numerical and experimental flow stress curve at $T=800^\circ C, \dot{\varepsilon}_{pl}=900s^{-1}$}
		\label{Bild:Abq_T800_900}
	\end{minipage}
\end{figure}

Figures \ref{Bild:Abq_T20_0_001} to \ref{Bild:Abq_T800_900} show graphical comparisons of the numerical prediction of the engineering stress-strain curve versus the corresponding experimental results for different temperatures and strain rates. 

The numerical stress strain curves follow qualitatively the experimental stress-strain curves but tend to be lower in general. In the quasi-static tensile test, figure \ref{Bild:JC20_0_001}, the assumption of adiabatic heating could be the cause for too low predicted yield strengths as the generated heat is not convected / conducted in the simulation and therefore leads to higher temperatures in the gauge length of the specimen. The largest deviation between experiment and simulation is for the test at $T=600^\circ C$ similar to the comparison in chapter \ref{Kap:AnalytischerVergleich}. As discussed in section \ref{Kap:ParameterMAnpassung} this behaviour was expected as the classic Johnson-Cook temperature term cannot describe the observed yield stress peak in this temperature region.

\clearpage

\section{Conclusions}
\label{Kap:Conclusions}

Material parameters for a Johnson-Cook flow stress model were derived based on quasi-static tensile tests as well as SHTB tests. The flow stress curve computed with this material parameter set is shown to be in a good agreement with the experiment within analytical and numerical comparisons. However, several improvements to the model are possible, but would require modifications to the Johnson-Cook flow stress model:
\begin{itemize}
	\item the first term of the Johnson-Cook flow stress could be replaced in order to improve the quasi-static yield curve, e.g. by a mixed Voce and Swift hardening term as for example used in \cite{Roth2016}
	\item the fit of the strain rate sensitivity C to the SHTB data in the range of strain rates of up to 1700/s is rather poor and could be better matched. In the tested strain rate range a linear description of the strain rate influence would suffice, but extrapolation to higher strain rates would presumably induce large errors. Testing at higher strain rates could give evidence but would require different tests and inverse identification methods, since SHTB procedures cannot reproduce such conditions. Since the strain rate sensitivity is low for this material, the overall error to the predicted yield stress is small
	\item The thermal softening is captured well, except for temperatures around $T=600^\circ C$. Using a modified temperature dependent term, as for example in \cite{Thimm2019} or \cite{Gerstgrasser2020}, could potentially capture this peak as well.% Doing so would require additional testing in an attempt to determine the temperature range where this yield stress peak occurs.
\end{itemize}
%
%% The Appendices part is started with the command \appendix;
%% appendix sections are then done as normal sections
%% \appendix

%% \section{}
%% \label{}

%% If you have bibdatabase file and want bibtex to generate the
%% bibitems, please use
%%
%%  \bibliographystyle{elsarticle-num} 
%%  \bibliography{<your bibdatabase>}

%% else use the following coding to input the bibitems directly in the
%% TeX file.

%\begin{thebibliography}{00}

%% \bibitem{label}
%% Text of bibliographic item

%\bibitem{}

%\end{thebibliography}

%\section{Acknowledgements}
%\label{Kap:Acknowledgements}
%
%\textcolor{red}{Wem gebührt hier Dank?}
%The authors would hereby like to thank the Swiss National Science Foundation (SNF) for the financial support under Grant No. \textcolor{red}{Evtl. SNF-Projekt von Darko?}.
%
\clearpage
\section*{References}
\bibliographystyle{plainnat}
\addcontentsline{toc}{section}{\refname}\bibliography{Literatur.bib}
\end{document}